\documentclass[structabstract]{aa}  
\usepackage{amsmath}
\usepackage{amssymb}

\usepackage{graphicx}
\usepackage{color}

\newcommand{\citep}[1]{(\cite{#1})} 

\newcommand{\HI}{$\mathrm{H_I}$ } 
\newcommand{\kb}{k_B}         
\newcommand{\Tsys}{T_{sys}}      

\newcommand{\TTlam}{ T_{21}(\vec{\Theta} ,\lambda)  } 
\newcommand{\TTlamz}{ T_{21}(\vec{\Theta} ,\lambda(z))  } 

\newcommand{\dlum}{d_L}
\newcommand{\dang}{d_A}

\newcommand{\hubb}{ h_{100} }    

\newcommand{\etaHI}{ n_{\tiny HI} }

\newcommand{\gHI}{ f_{H_I}}
\newcommand{\gHIz}{ f_{H_I}(z)}

\newcommand{\vis}{{\cal V}_{12} }
 
\newcommand{\LCDM}{$\Lambda \mathrm{CDM}$ }

\newcommand{\lgd}{\mathrm{log_{10}}}

\newcommand{\TrF}{\mathbf{T}}

\def\Mpc{\mathrm{Mpc}}
\def\hMpcm{\,h \,\Mpc^{-1}}

\def\kperp{k_\perp}
\def\kpar{k_\parallel}
\def\koperp{k_{BAO\perp }}
\def\kopar{k_{BAO\parallel}}

\usepackage{txfonts}
\begin{document}
   \title{21 cm observation of LSS at z $\sim$ 1 }

   \subtitle{Instrument sensitivity and foreground subtraction}

   \author{
          R. Ansari
          \inst{1} \inst{2}
          \and
          J.E. Campagne \inst{3}
         \and
          P.Colom  \inst{5} 
          \and
          J.M. Le Goff \inst{4}
          \and
          C. Magneville \inst{4}
          \and
          J.M. Martin  \inst{5} 
          \and
          M. Moniez \inst{3}
        \and 
          J.Rich \inst{4}
          \and 
          C.Y\`eche \inst{4}
          }

   \institute{
     Universit\'e Paris-Sud, LAL, UMR 8607, F-91898 Orsay Cedex, France 
   \and 
    CNRS/IN2P3,  F-91405 Orsay, France \\
     \email{ansari@lal.in2p3.fr}
    \and
    Laboratoire de lÍAcc\'el\'erateur Lin\'eaire, CNRS-IN2P3, Universit\'e Paris-Sud,
    B.P. 34, 91898 Orsay Cedex, France
    \and 
    CEA, DSM/IRFU, Centre d'Etudes de Saclay, F-91191 Gif-sur-Yvette, France 
    \and 
    GEPI, UMR 8111, Observatoire de Paris, 61 Ave de l'Observatoire, 75014 Paris, France
   }

   \date{Received July 15, 2011; accepted xxxx, 2011}

 
  \abstract
   { Large Scale Structures (LSS) in the universe can be traced using the neutral atomic hydrogen \HI through its 21 
cm emission. Such a 3D matter distribution map can be used to test the Cosmological model and to constrain the Dark Energy 
properties or its equation of state. A novel approach, called intensity mapping can be used to map the \HI distribution, 
using radio interferometers with large instanteneous field of view and waveband.}
  { In this paper, we study the sensitivity of different radio interferometer configurations, or multi-beam 
instruments for the observation of large scale structures and BAO oscillations in 21 cm and we discuss the problem of foreground removal. }
 { For each configuration, we determine instrument response by computing the (u,v) or Fourier angular frequency 
plane  coverage using visibilities. The (u,v) plane response is the noise power spectrum, 
hence the instrument sensitivity for LSS P(k) measurement. We describe also   a simple foreground subtraction method to 
separate LSS 21 cm signal from the foreground due to the galactic synchrotron and radio sources emission. }
   { We have computed the noise power spectrum for different instrument configuration as well as the extracted 
   LSS power spectrum, after separation of 21cm-LSS signal from the foregrounds. We have also obtained 
  the uncertainties on the Dark Energy parameters for an optimized 21 cm BAO survey.}
   { We show that a radio instrument with few hundred simultaneous beams and a collecting area of 
  $\sim 10000 \mathrm{m^2}$ will be able to  detect BAO signal at redshift z $\sim 1$ and will be 
  competitive with optical surveys. }


   \maketitle
%

\section{Introduction}

The study of the statistical properties of Large Scale Structure (LSS) in the Universe and their evolution 
with redshift is one the major tools in observational cosmology. These structures are usually mapped through 
optical observation of galaxies which are used as a tracer of the underlying matter distribution. 
An alternative and elegant approach for mapping the matter distribution, using neutral atomic hydrogen 
(\HI) as a tracer with intensity mapping, has been proposed in recent years \citep{peterson.06} \citep{chang.08}. 
Mapping the matter distribution using HI 21 cm emission as a tracer has been extensively discussed in literature
\citep{furlanetto.06} \citep{tegmark.09} and is being used in projects such as LOFAR \citep{rottgering.06} or
MWA \citep{bowman.07} to observe reionisation  at redshifts z $\sim$ 10. 

Evidence in favor of the acceleration of the expansion of the universe have been 
accumulated over the last twelve years, thanks to the observation of distant supernovae, 
CMB anisotropies and detailed analysis of the LSS.  
A cosmological Constant ($\Lambda$) or new cosmological 
energy density called {\em Dark Energy} has been advocated as the origin of this acceleration. 
Dark Energy is considered as one of the most intriguing puzzles in Physics and Cosmology. 
Several cosmological probes can be used to constrain the properties of this new cosmic fluid, 
more precisely its equation of state: The Hubble Diagram, or luminosity distance as a function 
of redshift of supernovae as standard candles, galaxy clusters, weak shear observations 
and Baryon Acoustic Oscillations (BAO). 

BAO are features imprinted  in the distribution of galaxies, due to the frozen 
sound waves which were present in the photon-baryon plasma prior to recombination 
at z $\sim$ 1100. 
This scale can be considered as a standard ruler with a comoving 
length  of $\sim 150 \mathrm{Mpc}$.
These features have been first observed in the CMB anisotropies
and are usually referred to as {\em acoustic peaks} (\cite{mauskopf.00}, \cite{larson.11}). 
The BAO modulation has been subsequently observed in the distribution of galaxies 
at low redshift ( $z < 1$) in the galaxy-galaxy correlation function by the SDSS 
\citep{eisenstein.05}  \citep{percival.07}  \citep{percival.10}, 2dGFRS  \citep{cole.05}  as well as 
WiggleZ \citep{blake.11} optical galaxy surveys.

Ongoing \citep{eisenstein.11}   or future surveys \citep{lsst.science}  
plan to measure precisely the BAO scale in the redshift range 
$0 \lesssim z \lesssim 3$, using either optical observation of galaxies  
or through 3D mapping Lyman $\alpha$ absorption lines toward distant quasars 
\citep{baolya},\citep{baolya2}. 
Radio observation of the 21 cm emission of neutral hydrogen appears as 
a very promising technique to map matter distribution up to redshift $z \sim 3$, 
complementary to optical surveys, especially in the optical redshift desert range 
$1 \lesssim z \lesssim 2$, and possibly up to the reionization redshift \citep{wyithe.08}.

In section 2, we  discuss the intensity mapping and its potential for measurement of the
\HI mass distribution power spectrum. The method used in this paper to characterize 
a radio instrument response and sensitivity for $P_{\mathrm{H_I}}(k)$ is presented in section 3.
We show also  the results for the 3D noise power spectrum for several  instrument configurations.
The contribution of foreground emissions due to the galactic synchrotron and radio sources
is described in section 4, as well as a simple component separation method. The performance of this 
method using two different sky models is also presented in section 4. 
The constraints which can be obtained on the Dark Energy parameters and DETF figure 
of merit for typical 21 cm intensity mapping survey are discussed in section 5.


\section{Intensity mapping and \HI power spectrum}


\subsection{21 cm intensity mapping}
Most of the cosmological information in the LSS is located at large scales
($ \gtrsim 1 \mathrm{deg}$), while the interpretation at smallest scales 
might suffer from the uncertainties on the non linear clustering effects.  
The BAO features in particular are at the degree angular scale on the sky 
and thus can be resolved easily with a rather modest size radio instrument 
(diameter $D \lesssim 100 \, \mathrm{m}$). The specific BAO clustering scale ($k_{\mathrm{BAO}}$) 
can be measured both in the transverse plane (angular correlation function, ($k_{\mathrm{BAO}}^\perp$) 
or along the longitudinal (line of sight or redshift ($k_{\mathrm{BAO}}^\parallel$) direction. A direct measurement of 
the Hubble parameter $H(z)$ can be obtained by comparing   the longitudinal and transverse 
BAO scales. A reasonably good redshift resolution $\delta z \lesssim 0.01$ is needed to resolve 
longitudinal BAO clustering, which is a challenge for photometric optical surveys.   

In order to obtain a measurement of the LSS power spectrum with small enough statistical
uncertainties (sample or cosmic variance),  a large volume of the universe should be observed, 
typically few $\mathrm{Gpc^3}$. Moreover, stringent constraint on DE parameters can only be 
obtained when comparing the distance or Hubble parameter measurements with 
DE models as a function of redshift, which requires a significant survey depth $\Delta z \gtrsim 1$.

Radio instruments intended for BAO surveys must thus have large instantaneous field
of view (FOV $\gtrsim 10 \, \mathrm{deg^2}$) and large bandwidth ($\Delta \nu \gtrsim 100 \, \mathrm{MHz}$)
to explore large redshift domains.

Although the application of 21 cm radio survey to cosmology, in particular LSS mapping has been 
discussed in length in the framework of large future instruments, such as the SKA (e.g \cite{ska.science}, \cite{abdalla.05}),
the method envisaged has been mostly through the detection of galaxies as \HI compact sources. 
However, extremely large radio telescopes are required to detected \HI sources at cosmological distances. 
The sensitivity (or detection threshold) limit $S_{lim}$ for the total power from the two polarisations
of a radio instrument characterized by an effective collecting area $A$, and system temperature $\Tsys$ can be written as 
\begin{equation}
S_{lim} = \frac{ \sqrt{2} \, \kb \, \Tsys }{ A \, \sqrt{t_{int} \delta \nu} } 
\end{equation} 
where $t_{int}$ is the total integration time and $\delta \nu$ is the detection frequency band. In table 
\ref{slims21} (left)  we have computed the sensitivity for 6 different sets of instrument effective area and system 
temperature, with a total integration time of 86400 seconds (1 day) over a frequency band of 1 MHz. 
The width of this frequency band is well adapted  to detection of \HI source with an intrinsic velocity 
dispersion of few 100 km/s. These detection limits should be compared with the expected 21 cm brightness
$S_{21}$ of compact sources which can be computed using the expression below (e.g.\cite{binney.98}) :
\begin{equation}
 S_{21}  \simeq  0.021 \mathrm{\mu Jy} \, \frac{M_{H_I} }{M_\odot}   \times 
\left( \frac{ 1\, \mathrm{Mpc}}{\dlum(z)} \right)^2 \times \frac{200 \, \mathrm{km/s}}{\sigma_v}  (1+z)
\end{equation}
 where $ M_{H_I} $ is the neutral hydrogen mass, $\dlum(z)$ is the luminosity distance and $\sigma_v$ 
is the source velocity dispersion. 
 
In table \ref{slims21} (right), we show the 21 cm brightness for 
compact objects with a total \HI \, mass of $10^{10} M_\odot$ and an intrinsic velocity dispersion of 
$200 \, \mathrm{km/s}$. The luminosity distance is computed for the standard 
WMAP \LCDM universe. $10^9 - 10^{10} M_\odot$ of neutral gas mass 
is typical for large galaxies \citep{lah.09}. It is clear that detection of \HI sources at cosmological distances
would require collecting area in the range of $10^6 \mathrm{m^2}$. 

Intensity mapping has been suggested as an alternative and economic method to map the 
3D distribution of neutral hydrogen by \citep{chang.08} and further studied by \citep{ansari.08} \citep{seo.10}. 
In this approach, sky brightness map with angular resolution $\sim 10-30 \, \mathrm{arc.min}$ is made for a 
wide range of frequencies. Each 3D pixel  (2 angles $\vec{\Theta}$, frequency $\nu$ or wavelength $\lambda$)  
would correspond to a cell with a volume of $\sim 10^3 \mathrm{Mpc^3}$, containing ten to hundred galaxies 
and a total \HI mass $ \sim 10^{12} M_\odot$. If we neglect local velocities relative to the Hubble flow, 
the observed frequency $\nu$ would be translated to the emission redshift $z$ through 
the well known relation:
\begin{eqnarray}
 z(\nu) & = & \frac{\nu_{21} -\nu}{\nu} 
\, ; \, \nu(z) = \frac{\nu_{21}}{(1+z)} 
\hspace{1mm} \mathrm{with}   \hspace{1mm}  \nu_{21} = 1420.4 \, \mathrm{MHz}  \\
 z(\lambda) & = & \frac{\lambda - \lambda_{21}}{\lambda_{21}} 
\, ; \, \lambda(z) = \lambda_{21} \times (1+z) 
\hspace{1mm} \mathrm{with}   \hspace{1mm}  \lambda_{21} = 0.211 \, \mathrm{m} 
\end{eqnarray}
The large scale distribution of the neutral hydrogen, down to angular scales of $\sim 10 \mathrm{arc.min}$ 
can then be observed without the detection of individual compact \HI sources, using the set of sky brightness 
map as a function of frequency (3D-brightness map) $B_{21}(\vec{\Theta},\lambda)$. The sky brightness $B_{21}$ 
(radiation power/unit solid angle/unit surface/unit frequency) 
can be converted to brightness temperature using the well known black body Rayleigh-Jeans approximation:
$$ B(T,\lambda) = \frac{ 2 \kb T }{\lambda^2} $$ 
 
\begin{table}
\begin{center}
\begin{tabular}{|c|c|c|}
\hline 
$A (\mathrm{m^2})$ & $ T_{sys} (K) $ & $ S_{lim} \, \mathrm{\mu Jy} $ \\
\hline 
5000 & 50 & 66 \\
5000 & 25 & 33 \\
100 000 & 50 & 3.3 \\
100 000 & 25 & 1.66 \\
500 000 & 50 & 0.66 \\
500 000 & 25 & 0.33 \\
\hline  
\end{tabular}
\hspace{3mm}
\begin{tabular}{|c|c|c|}
\hline
$z$ &  $\dlum \mathrm{(Mpc)}$ & $S_{21}  \mathrm{( \mu Jy)} $ \\
\hline      
0.25 &  1235   & 175 \\ 
0.50 &  2800   & 40   \\  
1.0   &  6600   & 9.6  \\  
1.5   &  10980 & 3.5  \\  
2.0   &  15710  & 2.5  \\ 
2.5  &  20690 &  1.7  \\  
\hline
\end{tabular}
\caption{Sensitivity or source detection limit for 1 day integration time (86400 s) and 1 MHz 
frequency band (left). Source 21 cm brightness for $10^{10} M_\odot$ \HI for different redshifts (right)  }
\label{slims21} 
\end{center}
\end{table}

\subsection{ \HI power spectrum and BAO}
In the absence of any foreground or background radiation, the brightness temperature 
for a given direction and wavelength $\TTlam$ would be proportional to 
the local \HI number density $\etaHI(\vec{\Theta},z)$ through the relation:
\begin{equation}
  \TTlamz  =   \frac{3}{32 \pi}  \, \frac{h}{\kb} \,  A_{21}  \, \lambda_{21}^2 \times
  \frac{c}{H(z)} \, (1+z)^2 \times  \etaHI (\vec{\Theta}, z)  
\end{equation}
where $A_{21}=2.85 \, 10^{-15} \mathrm{s^{-1}}$ \citep{astroformul} is the spontaneous 21 cm emission 
coefficient, $h$ is the Planck constant, $c$ the speed of light, $\kb$ the Boltzmann 
constant and $H(z)$ is the Hubble parameter at the emission redshift.
For a \LCDM universe and neglecting radiation energy density, the Hubble parameter
can be expressed as:
\begin{equation}
H(z)  \simeq  \hubb  \, \left[ \Omega_m (1+z)^3 + \Omega_\Lambda \right]^{\frac{1}{2}} 
\times  100 \, \, \mathrm{km/s/Mpc}  
\label{eq:expHz}
\end{equation}
Introducing the \HI mass fraction relative to the total baryon mass $\gHI$, the 
neutral hydrogen number density relative fluctuations can be written as, and the corresponding
21 cm emission temperature can be written as:
\begin{eqnarray}
\etaHI (\vec{\Theta}, z(\lambda) ) & = & \gHIz \times \Omega_B  \frac{\rho_{crit}}{m_{H}}  \times 
\left( \frac{\delta \rho_{H_I}}{\bar{\rho}_{H_I}} (\vec{\Theta},z) + 1 \right) \\
 \TTlamz  &  = & \bar{T}_{21}(z) \times \left( \frac{\delta \rho_{H_I}}{\bar{\rho}_{H_I}} (\vec{\Theta},z)  + 1 \right) 
\end{eqnarray}
where $\Omega_B, \rho_{crit}$ are respectively the present day mean baryon cosmological
and critical densities, $m_{H}$ is the hydrogen atom mass, and 
$\frac{\delta \rho_{H_I}}{\bar{\rho}_{H_I}}$ is the \HI density fluctuations. 

The present day neutral hydrogen fraction $\gHI(0)$ present in local galaxies has been 
measured to be $\sim 1\%$ of the baryon density \citep{zwann.05}:
$$ \Omega_{H_I} \simeq 3.5 \, 10^{-4} \sim 0.008 \times \Omega_B $$
The neutral hydrogen fraction is expected to increase with redshift, as gas is used 
in star formation during galaxy formation and evolution. Study of Lyman-$\alpha$ absorption 
indicate a factor 3 increase in the neutral hydrogen 
fraction at $z=1.5$ in the intergalactic medium \citep{wolf.05}, 
compared to its present day value $\gHI(z=1.5) \sim 0.025$. 
The 21 cm brightness temperature and the corresponding power spectrum can be written as 
(\cite{barkana.07} and \cite{madau.97}) :
\begin{eqnarray}
 P_{T_{21}}(k) & = & \left( \bar{T}_{21}(z)  \right)^2 \, P(k)    \label{eq:pk21z} \\
 \bar{T}_{21}(z)  & \simeq & 0.084  \, \mathrm{mK}  
\frac{ (1+z)^2 \, \hubb }{\sqrt{ \Omega_m (1+z)^3 + \Omega_\Lambda } } 
 \dfrac{\Omega_B}{0.044}  \,  \frac{\gHIz}{0.01} 
\label{eq:tbar21z}
\end{eqnarray}

\begin{figure}
\vspace*{-11mm}
\hspace{-5mm}
\includegraphics[width=0.57\textwidth]{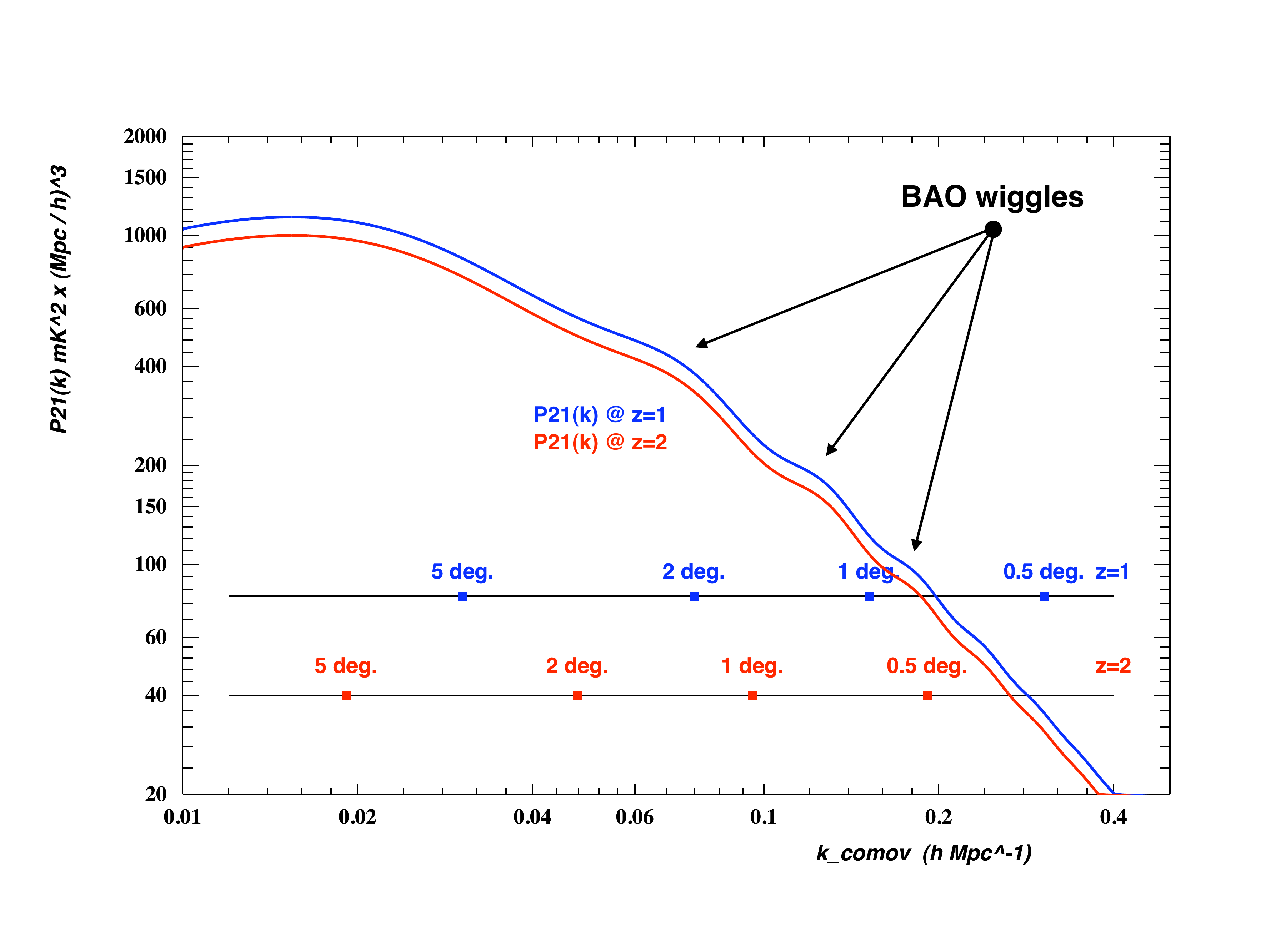}
\vspace*{-10mm}
\caption{\HI 21 cm emission power spectrum at redshifts z=1 (blue) and z=2 (red), with 
neutral gas fraction $\gHI=2\%$}
\label{figpk21}
\end{figure}

The table \ref{tabcct21} shows the mean 21 cm brightness temperature for the 
standard \LCDM cosmology and either a constant \HI mass fraction $\gHI = 0.01$, or 
linearly increasing  $\gHI \simeq 0.008 \times (1+z) $. Figure \ref{figpk21} shows the 
21 cm emission power spectrum at several redshifts, with a constant neutral fraction at 2\%
($\gHI=0.02$). The matter power spectrum has been computed using the 
\cite{eisenhu.98} parametrisation. The correspondence  with the angular scales is also 
shown for the standard WMAP \LCDM cosmology, according to the relation:
\begin{equation}
\theta_k = \frac{2 \pi}{k \, \dang(z) \, (1+z) }  
\hspace{3mm} 
k = \frac{2 \pi}{ \theta_k  \, \dang(z) \, (1+z) }  
\end{equation}
where $k$ is the comoving wave vector and $ \dang(z) $ is the angular diameter distance.

\begin{table}
\begin{center}
\begin{tabular}{|l|c|c|c|c|c|c|c|}
\hline 
\hline
 z  & 0.25 & 0.5 & 1. & 1.5 & 2. & 2.5 & 3. \\
\hline 
(a) $\bar{T}_{21}$ & 0.085 & 0.107 & 0.145 & 0.174 & 0.195 & 0.216 & 0.234 \\
\hline
(b) $\bar{T}_{21}$  & 0.085 & 0.128 & 0.232 & 0.348 & 0.468 & 0.605 & 0.749 \\
\hline
\hline  
\end{tabular}
\caption{Mean 21 cm brightness temperature in mK, as a function of redshift, for the 
standard \LCDM cosmology with constant \HI mass fraction at $\gHIz$=0.01  (a) or linearly 
increasing mass fraction (b)  $\gHIz=0.008(1+z)$ }
\label{tabcct21} 
\end{center}
\end{table}

\section{interferometric observations and P(k) measurement sensitivity }
\label{pkmessens}
\subsection{Instrument response}
\label{instrumresp}
We introduce briefly here the principles of interferometric observations and the definition of 
quantities useful for our calculations. Interested reader may refer to \citep{radastron} for a detailed 
and complete presentation of observation methods and signal processing in radio astronomy.  
In astronomy we are usually interested in measuring the sky emission intensity,
$I(\vec{\Theta},\lambda)$ in a given wave band, as a function of the sky direction. In radio astronomy
and interferometry in particular, receivers are sensitive to the sky emission complex 
amplitudes. However, for most sources, the phases vary randomly with a spatial correlation 
length significantly smaller than the instrument resolution.
\begin{eqnarray}
& & 
I(\vec{\Theta},\lambda)  =  | A(\vec{\Theta},\lambda) |^2  \hspace{2mm} , \hspace{1mm} I \in \mathbb{R}, A \in \mathbb{C} \\
& & < A(\vec{\Theta},\lambda) A^*(\vec{\Theta '},\lambda) >_{time}  = 0 \hspace{2mm} \mathrm{for}   \hspace{1mm} \vec{\Theta} \ne \vec{\Theta '}  
\end{eqnarray} 
A single receiver can be  characterized by its angular complex amplitude response $B(\vec{\Theta},\nu)$ and 
its position $\vec{r}$ in a reference frame. the waveform complex amplitude $s$ measured by the receiver, 
for each frequency can be written as a function of the electromagnetic wave vector 
$\vec{k}_{EM}(\vec{\Theta}, \lambda) $ :
\begin{equation}
s(\lambda)  =  \iint d \vec{\Theta} \, \, \, A(\vec{\Theta},\lambda) B(\vec{\Theta},\lambda) e^{i ( \vec{k}_{EM} . \vec{r} )} \\
\end{equation} 
We have set the electromagnetic (EM) phase origin at the center of the coordinate frame and 
the EM wave vector is related to the wavelength $\lambda$ through the usual equation
$ | \vec{k}_{EM} |  =  2 \pi / \lambda $. The receiver beam or antenna lobe $L(\vec{\Theta},\lambda)$ 
corresponds to the receiver intensity response:
\begin{equation}
L(\vec{\Theta}, \lambda) = B(\vec{\Theta},\lambda)  \,  B^*(\vec{\Theta},\lambda) 
\end{equation} 
The visibility signal of two receivers corresponds to the time averaged correlation between 
signals from two receivers. If we assume a sky signal with random uncorrelated phase, the 
visibility $\vis$ signal from two identical receivers, located at the position $\vec{r_1}$ and 
$\vec{r_2}$ can simply be written as a function of their position difference $\vec{\Delta r} = \vec{r_1}-\vec{r_2}$
\begin{equation}
\vis(\lambda) = < s_1(\lambda) s_2(\lambda)^* > = \iint d \vec{\Theta} \, \, I(\vec{\Theta},\lambda) L(\vec{\Theta},\lambda) 
e^{i ( \vec{k}_{EM} . \vec{\Delta r} ) }
\end{equation} 
This expression can be simplified if we consider receivers with narrow field of view 
($ L(\vec{\Theta},\lambda) \simeq  0$ for $| \vec{\Theta} | \gtrsim 10 \, \mathrm{deg.} $ ), 
and coplanar in respect to their common axis.
If we introduce two {\em Cartesian} like angular coordinates $(\alpha,\beta)$ centered at 
the common receivers axis, the visibilty would be written as the 2D Fourier transform
of the product of the sky intensity and the receiver beam, for the angular frequency 
\mbox{$(u,v)_{12} = 2 \pi( \frac{\Delta x}{\lambda} ,  \frac{\Delta y}{\lambda} )$}:
\begin{equation}
\vis(\lambda) \simeq  \iint d\alpha d\beta \, \, I(\alpha, \beta) \,  L(\alpha, \beta) 
\exp \left[ i 2 \pi \left( \alpha \frac{\Delta x}{\lambda} + \beta  \frac{\Delta y}{\lambda} \right) \right]
\end{equation} 
where $(\Delta x, \Delta y)$ are the two receiver distances on a plane perpendicular to 
the receiver axis. The $x$ and $y$ axis in the receiver plane are taken parallel to the 
two $(\alpha, \beta)$ angular planes. 

Furthermore, we introduce the conjugate Fourier variables $(u,v)$ and the Fourier transforms
of the sky intensity and the receiver beam:
\begin{center}
\begin{tabular}{ccc}
$(\alpha, \beta)$ & \hspace{2mm} $\longrightarrow $ \hspace{2mm} & $(u,v)$ \\
$I(\alpha, \beta, \lambda)$ & \hspace{2mm} $\longrightarrow $ \hspace{2mm} & ${\cal I}(u,v, \lambda)$ \\
$L(\alpha, \beta, \lambda)$ & \hspace{2mm} $\longrightarrow $ \hspace{2mm} & ${\cal L}(u,v, \lambda)$ \\
\end{tabular} 
\end{center}

The visibility can then be interpreted as the weighted sum of the sky intensity, in an angular 
wave number domain located around 
$(u, v)_{12}=2 \pi( \frac{\Delta x}{\lambda} ,  \frac{\Delta y}{\lambda} )$. The weight function is 
given by the receiver beam Fourier transform. 
\begin{equation}
\vis(\lambda) \simeq  \iint d u d v \, \, {\cal I}(u,v, \lambda) \, {\cal L}(u - 2 \pi \frac{\Delta x}{\lambda} , v - 2 \pi \frac{\Delta y}{\lambda} , \lambda)
\end{equation} 

A single receiver instrument would measure the total power integrated in a spot centered around the 
origin in the $(u,v)$ or the angular wave mode plane. The shape of the spot depends on the receiver 
beam pattern, but its extent would be $\sim 2 \pi D / \lambda$, where $D$ is the receiver physical 
size. 

\begin{figure}
\centering
\mbox{
\includegraphics[width=0.5\textwidth]{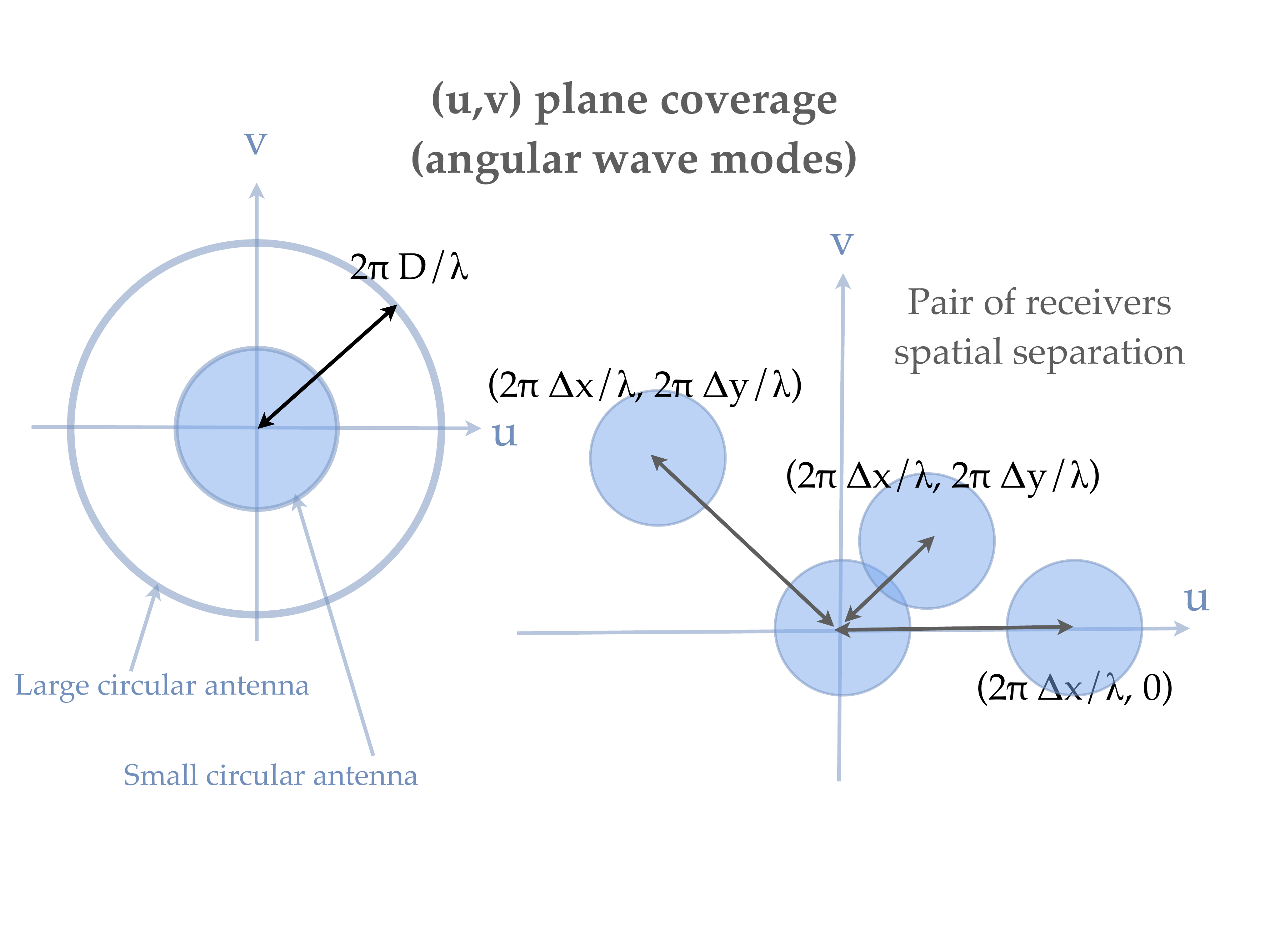}
}
\vspace*{-15mm}
\caption{Schematic view of the $(u,v)$ plane coverage by interferometric measurement.}
\label{figuvplane}
\end{figure}

The correlation signal from a pair of receivers would measure the integrated signal on a similar 
spot, located around the central angular wave mode  $(u, v)_{12}$ determined by the relative 
position of the two receivers (see figure \ref{figuvplane}).
In an interferometer with multiple receivers, the area covered by different receiver pairs in the 
$(u,v)$ plane might overlap and some pairs might measure the same area (same base lines). 
Several beams can be formed using different combination of the correlations from a set of  
antenna pairs.  

An instrument can thus be characterized by its $(u,v)$ plane coverage or response 
${\cal R}(u,v,\lambda)$. For a single dish with a single receiver in the focal plane, 
the instrument response is simply the Fourier transform of the beam. 
For a single dish with multiple receivers, either as a Focal Plane Array (FPA) or 
a multi-horn system, each beam (b) will have its own response 
${\cal R}_b(u,v,\lambda)$. 
For an interferometer, we can compute a raw instrument response 
${\cal R}_{raw}(u,v,\lambda)$ which corresponds to $(u,v)$ plane coverage by all 
receiver pairs  with uniform weighting. 
Obviously, different weighting schemes can be used, changing 
the effective beam shape and thus the response ${\cal R}_{w}(u,v,\lambda)$
and the noise behaviour. If the same Fourier angular frequency mode is measured 
by several receiver pairs, the raw instrument response might then be larger 
that unity. This non normalized instrument response is used to compute the projected 
noise power spectrum in the following section (\ref{instrumnoise}). 
We can also define a  normalized instrument response, ${\cal R}_{norm}(u,v,\lambda) \lesssim 1$ as:
\begin{equation}
{\cal R}_{norm}(u,v,\lambda) = {\cal R}(u,v,\lambda) / \mathrm{Max_{(u,v)}} \left[ {\cal R}(u,v,\lambda) \right] 
\end{equation} 
This normalized  instrument response can be used to compute the effective instrument beam, 
in particular in section \ref{recsec}.  

\subsection{Noise power spectrum}
\label{instrumnoise}
Let's consider a total power measurement using a receiver at wavelength $\lambda$, over a frequency 
bandwidth $\delta \nu$ centered on $\nu_0$, with an integration time $t_{int}$, characterized by a system temperature 
$\Tsys$. The uncertainty or fluctuations of this measurement due to the receiver noise can be written as 
$\sigma_{noise}^2 = \frac{2 \Tsys^2}{t_{int} \, \delta \nu}$. This term 
corresponds also to the noise for the visibility $\vis$ measured from two identical receivers, with uncorrelated 
noise. If the receiver has an effective area $A \simeq \pi D^2/4$ or $A \simeq D_x D_y$, the measurement 
corresponds to the integration of power over a spot in the angular frequency plane with an area $\sim A/\lambda^2$. The noise spectral density, in the angular frequencies plane (per unit area of angular frequencies  $\frac{\delta u}{ 2 \pi} \times \frac{\delta v}{2 \pi}$), corresponding to a visibility 
measurement from a pair of receivers can be written as:
\begin{eqnarray}
P_{noise}^{\mathrm{pair}} & = & \frac{\sigma_{noise}^2}{ A / \lambda^2 }  \\
P_{noise}^{\mathrm{pair}} & \simeq & \frac{2 \, \Tsys^2 }{t_{int}  \, \delta \nu} \, \frac{ \lambda^2 }{ D^2 } 
\hspace{5mm} \mathrm{units:} \, \mathrm{K^2 \times rad^2} 
\label{eq:pnoisepairD}
\end{eqnarray}

The sky temperature measurement can thus be characterized by the noise spectral power density in 
the angular frequencies plane $P_{noise}^{(u,v)} \simeq \frac{\sigma_{noise}^2}{A / \lambda^2}$, in $\mathrm{Kelvin^2}$  
per unit area of angular frequencies  $\frac{\delta u}{ 2 \pi} \times \frac{\delta v}{2 \pi}$:
We can characterize the sky temperature measurement with a radio instrument by the noise 
spectral power density in the angular frequencies plane $P_{noise}(u,v)$ in units of $\mathrm{Kelvin^2}$  
per unit area of angular frequencies  $\frac{\delta u}{ 2 \pi} \times \frac{\delta v}{2 \pi}$. 
For an interferometer made of identical receiver elements, several ($n$) receiver pairs 
might have the same baseline. The noise power density in the corresponding $(u,v)$ plane area 
is then reduced by a factor $1/n$. More generally, we can write the instrument  noise 
spectral power density using the instrument response defined in section \ref{instrumresp} :
\begin{equation}
P_{noise}(u,v) = \frac{ P_{noise}^{\mathrm{pair}} } { {\cal R}_{raw}(u,v,\lambda) } 
\end{equation} 

When the intensity maps are projected in a three dimensional box in the universe and the 3D power spectrum 
$P(k)$ is computed, angles are translated into comoving transverse distances, 
and frequencies or wavelengths into comoving radial distance, using the following relations:
\begin{eqnarray}
\delta \alpha , \beta & \rightarrow & \delta \ell_\perp = (1+z) \, \dang(z) \, \delta \alpha,\beta  \\
\delta \nu & \rightarrow & \delta \ell_\parallel = (1+z) \frac{c}{H(z)} \frac{\delta \nu}{\nu} 
  = (1+z) \frac{\lambda}{H(z)} \delta \nu \\
\delta u , \delta v & \rightarrow & \delta k_\perp = \frac{ \delta u \, , \, \delta v }{  (1+z) \, \dang(z)  } \\
\frac{1}{\delta \nu} & \rightarrow & \delta k_\parallel = \frac{H(z)}{c} \frac{1}{(1+z)} \, \frac{\nu}{\delta \nu}
 =  \frac{H(z)}{c} \frac{1}{(1+z)^2} \, \frac{\nu_{21}}{\delta \nu}
\end{eqnarray}

If we consider a uniform noise spectral density in the $(u,v)$ plane corresponding to the 
equation \ref{eq:pnoisepairD} above,  the three dimensional projected noise spectral density 
can then be written as: 
\begin{equation}
P_{noise}(k) = 2 \, \frac{\Tsys^2}{t_{int} \, \nu_{21} } \, \frac{\lambda^2}{D^2}  \, \dang^2(z) \frac{c}{H(z)} \, (1+z)^4  
\label{ctepnoisek}
\end{equation} 

$P_{noise}(k)$ would be in units of $\mathrm{mK^2 \, Mpc^3}$ with $\Tsys$ expressed in $\mathrm{mK}$, 
$t_{int}$ is the integration time expressed in second, 
$\nu_{21}$ in $\mathrm{Hz}$, $c$ in $\mathrm{km/s}$, $\dang$ in $\mathrm{Mpc}$ and 
 $H(z)$ in $\mathrm{km/s/Mpc}$. 

The matter or \HI distribution power spectrum determination statistical errors vary as the number of 
observed Fourier modes, which is inversely proportional to volume of the universe 
which is observed (sample variance).  As the observed volume is proportional to the 
surveyed solid angle, we  consider the survey of a fixed
fraction of the sky, defined by  total solid angle $\Omega_{tot}$, performed during a determined 
total observation time $t_{obs}$. 
A single dish instrument with diameter $D$ would have an instantaneous field of view 
$\Omega_{FOV} \sim \left( \frac{\lambda}{D} \right)^2$, and would require 
a number of pointings  $N_{point} = \frac{\Omega_{tot}}{\Omega_{FOV}}$ to cover the survey area. 
Each sky direction or pixel of size $\Omega_{FOV}$ will be observed during an integration 
time $t_{int} = t_{obs}/N_{point} $. Using equation \ref{ctepnoisek} and the previous expression
for the integration time, we can compute a simple expression
for the noise spectral power density by a single dish instrument of diameter $D$:
\begin{equation}
P_{noise}^{survey}(k) = 2 \, \frac{\Tsys^2 \, \Omega_{tot} }{t_{obs} \, \nu_{21} } \, \dang^2(z) \frac{c}{H(z)} \, (1+z)^4  
\end{equation}

It is important to note that any real instrument do not have a flat 
response in the $(u,v)$ plane, and the observations provide no information above 
a certain maximum angular frequency $u_{max},v_{max}$. 
One has to take into account either a damping of the observed sky power 
spectrum or an increase of the noise spectral power if 
the observed power spectrum is corrected for damping. The white noise 
expressions given below should thus be considered as a lower limit or floor of the 
instrument noise spectral density. 
 
For a single dish instrument of diameter $D$ equipped with a multi-feed or 
phase array receiver system, with $N$ independent beams on sky, 
the noise spectral density decreases by a factor $N$, 
thanks to the  increase of per pointing integration time:

\begin{equation}
P_{noise}^{survey}(k) = \frac{2}{N} \, \frac{\Tsys^2 \, \Omega_{tot} }{t_{obs} \, \nu_{21} } \, \dang^2(z) \frac{c}{H(z)} \, (1+z)^4  
\label{eq:pnoiseNbeam}
\end{equation}

This expression (eq. \ref{eq:pnoiseNbeam}) can also be used for a filled interferometric array of $N$ 
identical receivers with a  total collection area $\sim D^2$. Such an array could be made for example 
of $N=q \times q$ {\it small dishes}, each with diameter $D/q$, arranged as $q \times q$ square.   

For a single dish of diameter $D$, or an interferometric instrument with maximal extent $D$, 
observations provide information up to $u_{max},v_{max} \lesssim 2 \pi D / \lambda $. This value of 
$u_{max},v_{max}$ would be mapped to a maximum transverse cosmological wave number
$k^{\perp}_{max}$:
\begin{equation}
k^{\perp}  =  \frac{(u,v)}{(1+z) \dang}  \hspace{8mm} 
k^{\perp}_{max}  \lesssim  \frac{2 \pi}{\dang \, (1+z)^2} \frac{D}{\lambda_{21}} 
\label{kperpmax}
\end{equation}   

\begin{figure}
\vspace*{-25mm}
\centering
\mbox{
\hspace*{-10mm}
\includegraphics[width=0.65\textwidth]{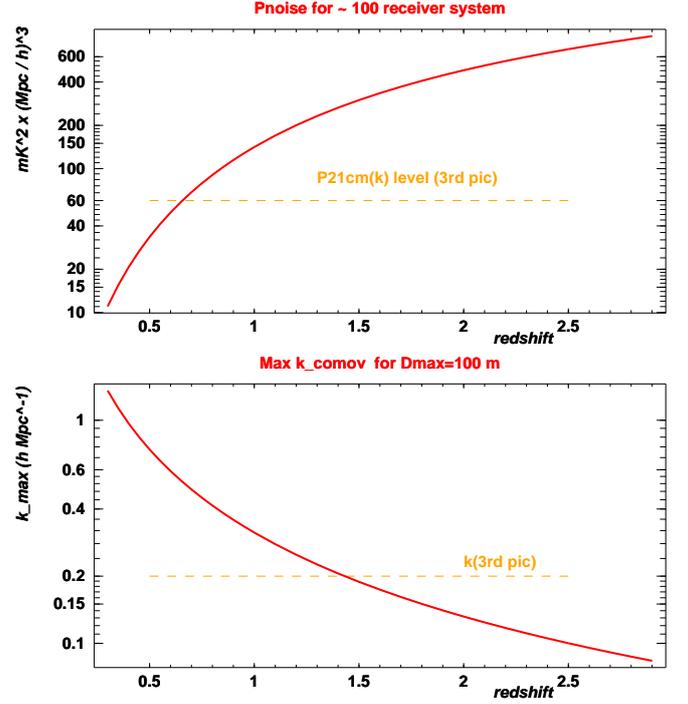}
}
\vspace*{-40mm}
\caption{Minimal noise level for a 100 beams instrument with \mbox{$\Tsys=50 \mathrm{K}$} 
as a function of redshift (top).  Maximum $k$ value for  a 100 meter diameter primary antenna (bottom) }
\label{pnkmaxfz}
\end{figure}

Figure \ref{pnkmaxfz} shows the evolution of the noise spectral density $P_{noise}^{survey}(k)$
as a function of redshift, for a radio survey of the sky, using an instrument with $N=100$
beams and a system noise temperature $\Tsys = 50 \mathrm{K}$.
The survey is supposed to cover a quarter of sky $\Omega_{tot} = \pi \, \mathrm{srad}$, in one 
year. The maximum comoving wave number $k_{max}$  is also shown as a function 
of redshift, for an instrument with $D=100 \, \mathrm{m}$ maximum extent. In order 
to take into account the radial component of $\vec{k}$ and the increase of 
the instrument noise level with $k^{\perp}$, we have taken the effective $k_{ max}  $ 
as half of the maximum transverse $k^{\perp} _{max}$ of \mbox{eq. \ref{kperpmax}}: 
\begin{equation}
k_{max} (z) = \frac{\pi}{\dang \, (1+z)^2} \frac{D=100 \mathrm{m}}{\lambda_{21}} 
\end{equation}

\subsection{Instrument configurations and noise power spectrum}
\label{instrumnoise}
\begin{figure}
\centering
\vspace*{-15mm}
\mbox{
\hspace*{-10mm}
\includegraphics[width=0.5\textwidth]{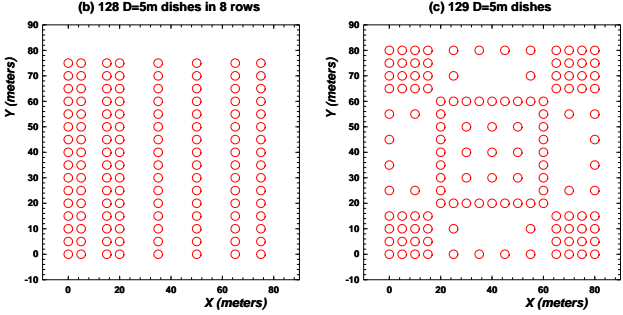}
}
\vspace*{-15mm}
\caption{ Array layout for configurations (b) and (c) with 128 and 129  D=5 meter 
diameter dishes. }
\label{figconfbc}
\end{figure}

We have numerically computed the instrument response ${\cal R}(u,v,\lambda)$ 
with uniform weights in the $(u,v)$ plane for several instrument configurations:
\begin{itemize}
\item[{\bf a} :] A packed array of $n=121 \, D_{dish}=5 \, \mathrm{m}$ dishes, arranged in 
a square $11 \times 11$ configuration ($q=11$). This array covers an area of 
$55 \times 55 \, \mathrm{m^2}$ 
\item [{\bf b} :] An array of $n=128  \, D_{dish}=5 \, \mathrm{m}$ dishes, arranged 
in 8 rows, each with 16 dishes. These 128 dishes are spread over an area 
$80 \times 80  \, \mathrm{m^2}$. The array layout for this configuration is
shown in figure \ref{figconfbc}. 
\item [{\bf c} :] An array of $n=129  \, D_{dish}=5 \, \mathrm{m}$ dishes, arranged 
 over an area $80 \times 80  \, \mathrm{m^2}$. This configuration has in 
particular 4 sub-arrays of packed 16 dishes ($4\times4$), located in the 
four array corners. This array layout is also shown figure \ref{figconfbc}. 
\item [{\bf d} :] A single dish instrument, with diameter $D=75 \, \mathrm{m}$, 
equipped with a 100 beam focal plane receiver array. 
\item[{\bf e} :] A packed array of $n=400 \, D_{dish}=5 \, \mathrm{m}$ dishes, arranged in 
a square $20 \times 20$ configuration ($q=20$). This array covers an area of 
$100 \times 100 \, \mathrm{m^2}$ 
\item[{\bf f} :] A packed array of 4 cylindrical reflectors, each 85 meter long and 12 meter 
wide. The focal line of each cylinder is equipped with 100 receivers, each 
$2 \lambda$ long, corresponding to $\sim 0.85 \, \mathrm{m}$ at $z=1$. 
This array covers an area of $48 \times 85 \, \mathrm{m^2}$, and have 
a total of $400$ receivers per polarisation, as in the (e) configuration.
We have computed the noise power spectrum for {\em perfect} 
cylinders, where all receiver pair correlations are used (fp), or for 
a non perfect instrument, where only correlations between receivers 
from different cylinders are used.
\item[{\bf g} :] A packed array of 8 cylindrical reflectors, each 102 meter long and 12 meter 
wide. The focal line of each cylinder is equipped with 120 receivers, each 
$2 \lambda$ long, corresponding to $\sim 0.85 \, \mathrm{m}$ at $z=1$. 
This array covers an area of $96 \times 102 \, \mathrm{m^2}$ and has 
a total of 960  receivers per polarisation. As for the (f) configuration,  
we have computed the noise power spectrum for {\em perfect} 
cylinders, where all receiver pair correlations are used (gp), or for 
a non perfect instrument, where only correlations between receivers 
from different cylinders are used.
\end{itemize}

We have used simple triangular shaped dish response in the $(u,v)$ plane. 
However, we have introduced a filling factor or illumination efficiency 
$\eta$, relating the effective dish diameter $D_{ill}$ to the 
mechanical dish size $D^{ill} = \eta \, D_{dish}$. The effective area $A_e \propto \eta^2$ scales 
as $\eta^2$ or $\eta_x \eta_y$. 
\begin{eqnarray}
{\cal L}_\circ (u,v,\lambda) & = & \bigwedge_{[\pm 2 \pi D^{ill}/ \lambda]}(\sqrt{u^2+v^2})  \\
 L_\circ (\alpha,\beta,\lambda) & = & \left[ \frac{ \sin (\pi (D^{ill}/\lambda) \sin \theta ) }{\pi (D^{ill}/\lambda) \sin \theta} \right]^2 
\hspace{4mm} \theta=\sqrt{\alpha^2+\beta^2}
\end{eqnarray}
For the multi-dish configuration studied here, we have taken the illumination efficiency factor
{\bf $\eta = 0.9$}. 

\begin{figure*}
\centering
\mbox{
\hspace*{-10mm}
\includegraphics[width=0.90\textwidth]{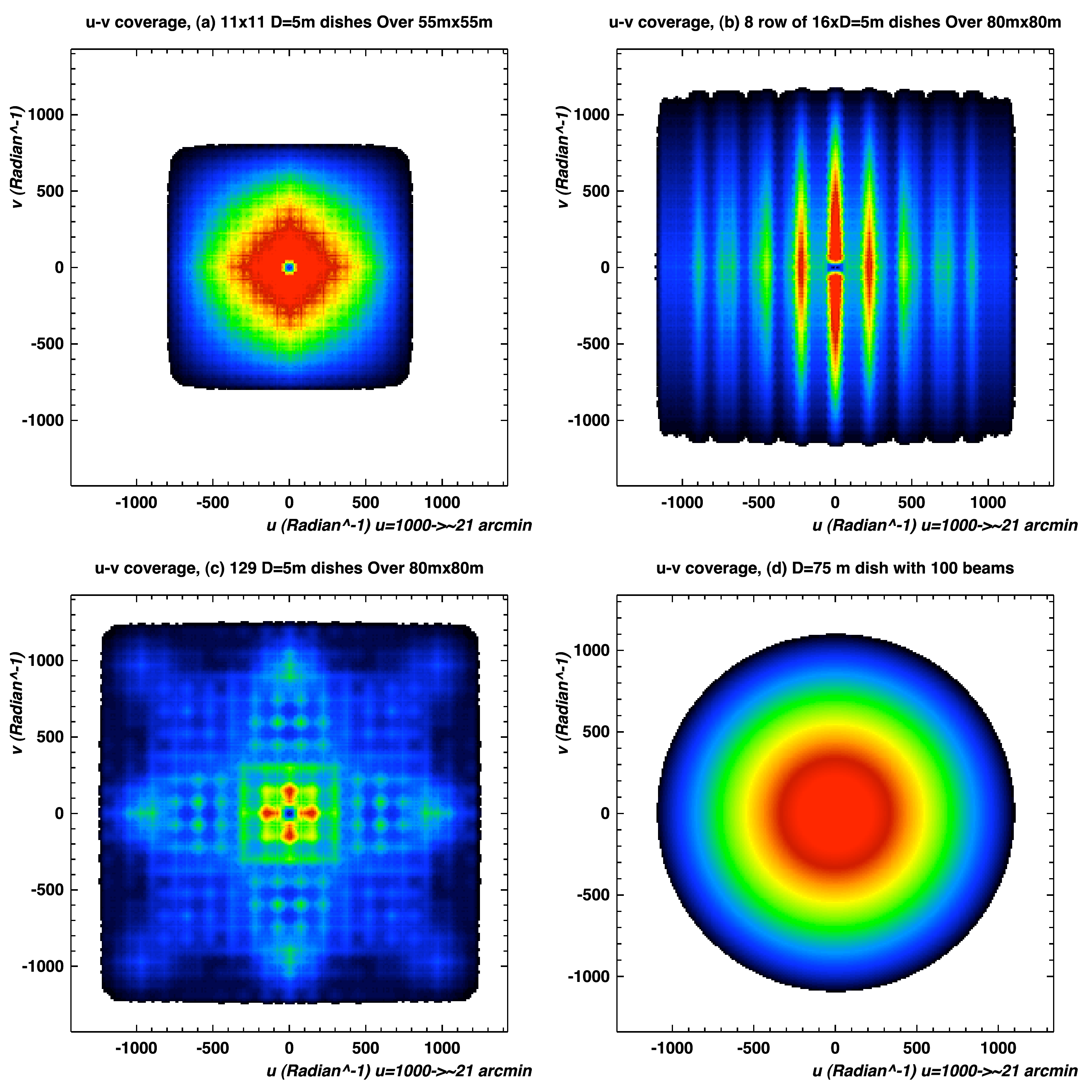}
}
\caption{(u,v) plane coverage (raw instrument response ${\cal R}(u,v,\lambda)$ 
for four configurations.
(a) 121 $D_{dish}=5$ meter diameter dishes arranged in a compact, square array 
of $11 \times 11$, (b) 128 dishes arranged in 8 row of 16 dishes each (fig. \ref{figconfbc}), 
(c) 129 dishes arranged as shown in figure \ref{figconfbc} , (d) single D=75 meter diameter, with 100 beams. 
(color scale : black $<1$, blue, green, yellow, red $\gtrsim 80$) }
\label{figuvcovabcd}
\end{figure*}

\begin{figure*}
\vspace*{-25mm}
\centering
\mbox{
\hspace*{-20mm}
\includegraphics[width=1.15\textwidth]{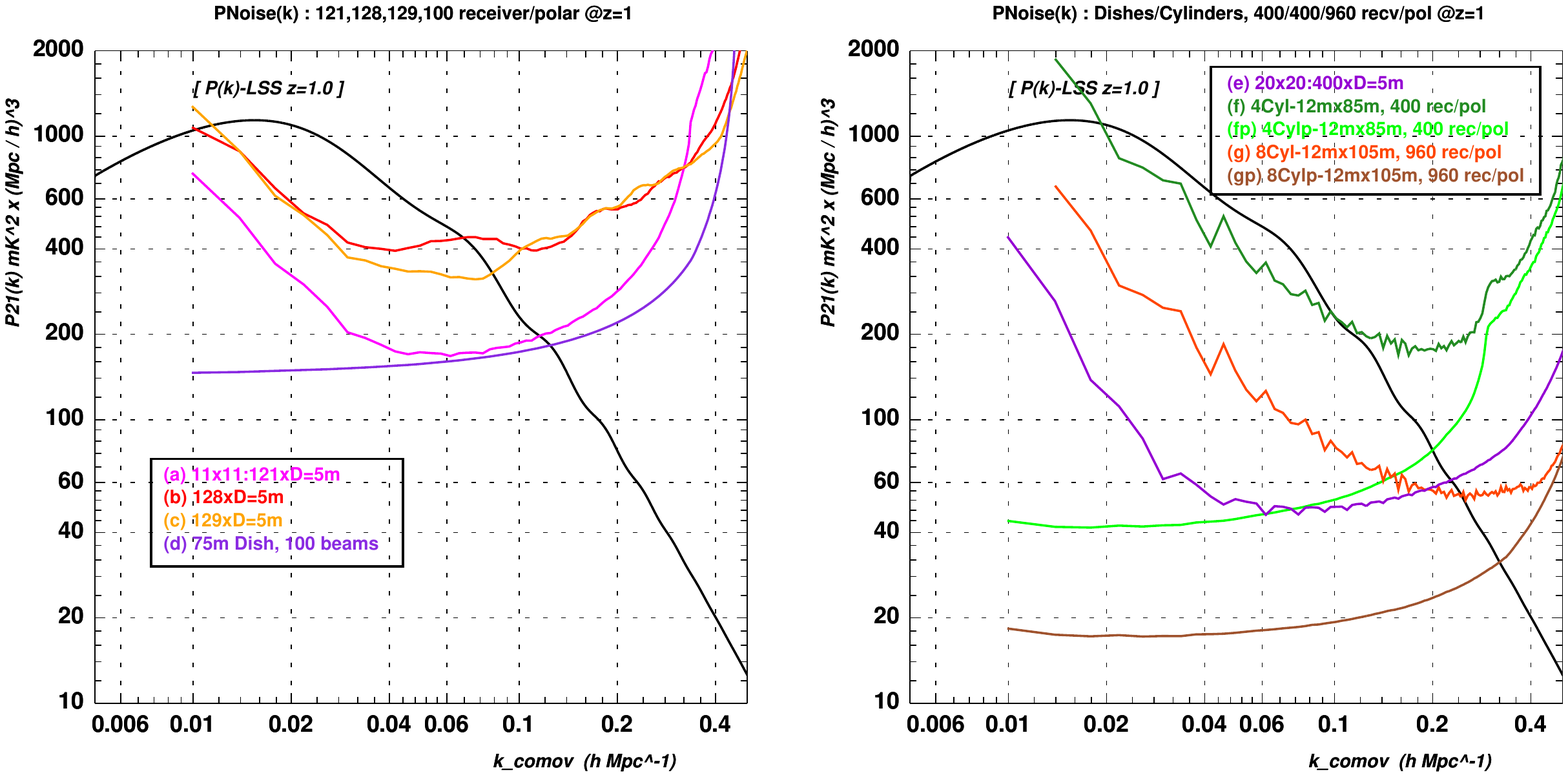}
}
\vspace*{-40mm}
\caption{P(k) LSS power  and noise power spectrum for several interferometer 
configurations ((a),(b),(c),(d),(e),(f),(g)) with 121, 128, 129, 400 and 960 receivers.}
\label{figpnoisea2g}
\end{figure*}

For the receivers along the focal line of cylinders, we have assumed that the 
individual receiver response in the $(u,v)$ plane corresponds to one from a 
rectangular shaped antenna. The illumination efficiency factor has been taken 
equal to $\eta_x = 0.9$ in the direction of the cylinder width, and $\eta_y = 0.8$ 
along the cylinder length. It should be noted that the small angle approximation 
used here for the expression of visibilities is not valid for the receivers along 
the cylinder axis. However, some preliminary numerical checks indicate that 
the results obtained here for the noise spectral power density  would not change significantly.
The instrument responses shown here correspond to fixed pointing toward the zenith, which 
is the case for a transit type telescope.

\begin{equation}
 {\cal L}_\Box(u,v,\lambda)  = 
\bigwedge_{[\pm 2 \pi D^{ill}_x / \lambda]} (u ) \times
\bigwedge_{[\pm 2 \pi D^{ill}_y / \lambda ]} (v ) 
\end{equation}
Figure \ref{figuvcovabcd} shows the instrument response ${\cal R}(u,v,\lambda)$ 
for the four configurations (a,b,c,d) with $\sim 100$ receivers per 
polarisation. The resulting projected noise spectral power density is shown in figure 
\ref{figpnoisea2g}. The increase of $P_{noise}(k)$ at low $k^{comov} \lesssim 0.02$ 
is due to the fact that we have ignored all auto-correlation measurements.  
It can be seen that an instrument with $100-200$ beams and $\Tsys = 50 \mathrm{K}$ 
should have enough sensitivity to map LSS in 21 cm at redshift z=1.

\section{ Foregrounds and Component separation }
\label{foregroundcompsep}
Reaching the required sensitivities is not the only difficulty of observing the large 
scale structures in 21 cm. Indeed, the synchrotron emission of the 
Milky Way and the extra galactic radio sources are a thousand times brighter than the 
emission of the neutral hydrogen distributed in the universe. Extracting the LSS signal
using Intensity Mapping, without identifying the \HI point sources is the main challenge 
for this novel observation method. Although this task might seem impossible at first, 
it has been suggested that the smooth frequency dependence of the synchrotron
emissions can be used to separate the faint LSS signal from the Galactic and radio source 
emissions.
However, any real radio instrument has a beam shape which changes with 
frequency: this instrumental effect significantly increases the difficulty and complexity of this component separation 
technique. The effect of frequency dependent beam shape is some time referred to as {\em 
mode mixing}. See for example \citep{morales.06}, \citep{bowman.07}. 

In this section, we present a short description of the foreground emissions and 
the simple models we have used for computing the sky radio emissions in the GHz frequency 
range. We present also a simple component separation method to extract the LSS signal and 
its performance. We show in particular the effect of the instrument response on the recovered 
power spectrum. The results presented in this section concern the 
total sky emission and the LSS 21 cm signal extraction in the $z \sim 0.6$ redshift range,
corresponding to the central frequency $\nu \sim 884$ MHz.  
 
\subsection{ Synchrotron and radio sources }
We have modeled the radio sky in the form of three dimensional maps (data cubes) of sky temperature 
brightness $T(\alpha, \delta, \nu)$ as a function of two equatorial angular coordinates $(\alpha, \delta)$ 
and the frequency $\nu$. Unless otherwise specified, the results presented here are based on simulations of 
$90 \times 30 \simeq 2500 \, \mathrm{deg^2}$ of the sky, centered on $\alpha= 10\mathrm{h}00\mathrm{m} , \delta=+10 \, \mathrm{deg.}$, and  covering 128 MHz in frequency. We have selected this particular area of the sky  in order to minimize 
the Galactic synchrotron foreground. The sky cube characteristics (coordinate range, size, resolution) 
used in the simulations are given in the table \ref{skycubechars}.
\begin{table}
\begin{center}
\begin{tabular}{|c|c|c|}
\hline 
 & range & center  \\
\hline 
Right ascension & 105 $ < \alpha < $ 195 deg. &  150 deg.\\
Declination & -5 $ < \delta < $ 25 deg. & +10 deg. \\
Frequency & 820 $ < \nu < $ 948 MHz & 884 MHz \\
Wavelength & 36.6 $ < \lambda < $ 31.6 cm & 33.9 cm \\
Redshift & 0.73 $ < z < $ 0.5 & 0.61 \\
\hline 
\hline
& resolution & N-cells \\
\hline
Right ascension & 3 arcmin & 1800 \\
Declination & 3 arcmin & 600 \\
Frequency & 500 kHz ($d z \sim 10^{-3}$) & 256 \\
\hline
\end{tabular} \\[1mm]
\end{center}
\caption{
Sky cube characteristics for the simulation performed in this paper. 
Cube size : $ 90 \, \mathrm{deg.} \times 30 \, \mathrm{deg.} \times 128 \, \mathrm{MHz}$   
$ 1800 \times 600 \times 256 \simeq 123 \, 10^6$ cells 
}
\label{skycubechars}
\end{table}
\par 
Two different methods have been used to compute the sky temperature data cubes.
We have used the Global Sky Model (GSM) \citep{gsm.08} tools to generate full sky 
maps of the emission temperature at different frequencies, from which we have 
extracted the brightness temperature cube for the region defined above 
(Model-I/GSM $T_{gsm}(\alpha, \delta, \nu)$).
As the GSM maps have an intrinsic resolution of $\sim$ 0.5 degree, it is 
difficult to have reliable results for the effect of point sources on the reconstructed 
LSS power spectrum.

We have thus made also a simple sky model using the Haslam Galactic synchrotron map
at 408 MHz \citep{haslam.82} and the NRAO VLA Sky Survey (NVSS) 1.4 GHz radio source 
catalog \citep{nvss.98}. The sky temperature cube in this model (Model-II/Haslam+NVSS) 
has been computed through the following steps:

\begin{enumerate}
\item The Galactic synchrotron emission is modeled as a power law with spatially 
varying spectral index. We assign a power law index $\beta = -2.8  \pm 0.15$ to each sky direction.
$\beta$ has a gaussian distribution centered at -2.8 and with standard 
deviation $\sigma_\beta = 0.15$.
The synchrotron contribution to the sky temperature for each cell is then 
obtained  through the formula:
$$ T_{sync}(\alpha, \delta, \nu) = T_{haslam} \times \left(\frac{\nu}{408 \, \mathrm{MHz}}\right)^\beta $$
\item A two dimensional $T_{nvss}(\alpha,\delta)$ sky brightness temperature at 1.4 GHz is computed 
by projecting the radio sources in the NVSS catalog to a grid with the same angular resolution as 
the sky cubes. The source brightness in Jansky is converted to temperature taking the 
pixel angular size into account ($ \sim 21 \mathrm{mK / mJansky}$ at 1.4 GHz and $3' \times 3'$ pixels).  
A spectral index $\beta_{src} \in [-1.5,-2]$ is also assigned to each sky direction for the radio source 
map; we have taken $\beta_{src}$ as a flat random number in the range $[-1.5,-2]$, and the 
contribution of the radiosources to the sky temperature is computed as follows:
$$ T_{radsrc}(\alpha, \delta, \nu) = T_{nvss} \times \left(\frac{\nu}{1420 \, \mathrm{MHz}}\right)^{\beta_{src}} $$
\item The sky brightness temperature data cube is obtained through the sum of
the two contributions, Galactic synchrotron and resolved radio sources: 
$$ T_{fgnd}(\alpha, \delta, \nu) = T_{sync}(\alpha, \delta, \nu) + T_{radsrc}(\alpha, \delta, \nu) $$ 
\end{enumerate}

 The 21 cm temperature fluctuations due to neutral hydrogen in large scale structures
$T_{lss}(\alpha, \delta, \nu)$  have been computed using the 
SimLSS \footnote{SimLSS : {\tt http://www.sophya.org/SimLSS} }  software package: 
complex normal Gaussian fields were first generated in Fourier space.
The amplitude of each mode was then multiplied by the square root
of the power spectrum $P(k)$ at $z=0$ computed according to the parametrization of 
\citep{eisenhu.98}. We have used the standard cosmological parameters, 
 $H_0=71 \, \mathrm{km/s/Mpc}$, $\Omega_m=0.27$, $\Omega_b=0.044$,
$\Omega_\lambda=0.73$ and $w=-1$.
An inverse FFT was then performed to compute the matter density fluctuations $\delta \rho / \rho$ 
in the linear regime, 
in a box of $3420 \times 1140 \times 716  \, \mathrm{Mpc^3}$ and evolved
to redshift $z=0.6$. 
The size of the box is about 2500 $\mathrm{deg^2}$  in the transverse direction and
$\Delta z \simeq 0.23$ in the longitudinal direction.  
The size of the cells is  $1.9 \times 1.9 \times 2.8 \, \mathrm{Mpc^3}$, which correspond approximately to the 
sky cube angular and frequency resolution defined above.  

The mass fluctuations has been 
converted into temperature through a factor $0.13 \, \mathrm{mK}$, corresponding to a hydrogen 
fraction $0.008 \times (1+0.6)$, using equation \ref{eq:tbar21z}.  
The total sky brightness temperature is then computed as the sum 
of foregrounds and the LSS 21 cm emission:
$$  T_{sky} = T_{sync}+T_{radsrc}+T_{lss}   \hspace{5mm} OR \hspace{5mm} 
T_{sky} = T_{gsm}+T_{lss} $$ 

\begin{figure}
\centering
\vspace*{-10mm}
\mbox{
\hspace*{-20mm}
\includegraphics[width=0.6\textwidth]{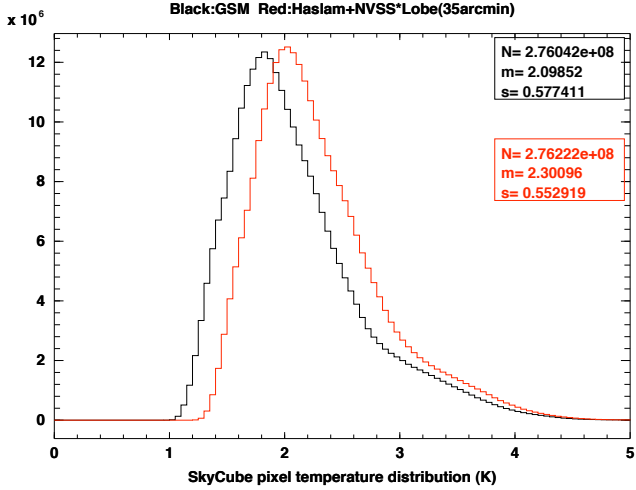}
}
\vspace*{-10mm}
\caption{Comparison of GSM (black) Model-II (red) sky cube temperature distribution.
The Model-II (Haslam+NVSS),
has been smoothed with a 35 arcmin gaussian beam. }
\label{compgsmhtemp}
\end{figure}

\begin{figure*}
\centering
\mbox{
\includegraphics[width=0.9\textwidth]{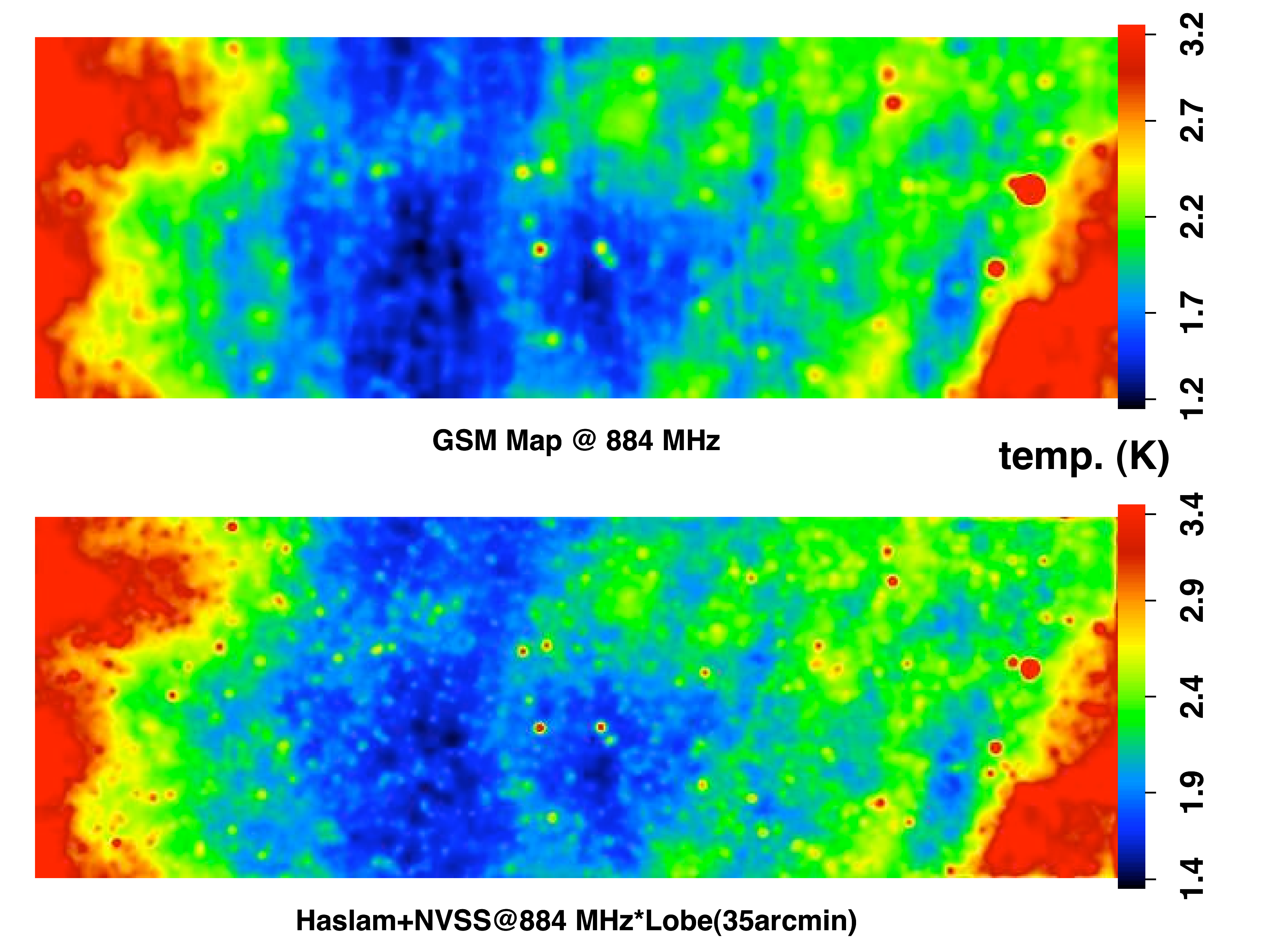}
}
\caption{Comparison of GSM map (top) and Model-II sky map at 884 MHz (bottom). 
The Model-II (Haslam+NVSS) has been smoothed with a 35 arcmin (FWHM) gaussian beam.}
\label{compgsmmap}
\end{figure*}

Table \ref{sigtsky} summarizes the mean and standard deviation of the sky brightness 
temperature $T(\alpha, \delta, \nu)$ for the different components computed in this study.
It should be noted that the standard deviation depends on the map resolution and the values given 
in table \ref{sigtsky}  correspond to sky cubes computed here, with $\sim 3$ arc minute
angular and 500 kHz frequency resolutions (see table \ref{skycubechars}).
Figure \ref{compgsmmap} shows the comparison of the GSM temperature map at 884 MHz 
with Haslam+NVSS map, smoothed with a 35 arcmin gaussian beam. 
Figure \ref{compgsmhtemp} shows the comparison of the sky cube temperature distribution 
for Model-I/GSM and Model-II. There is good agreement between the two models, although 
the mean temperature for Model-II is slightly higher ($\sim 10\%$) than Model-I.

\begin{figure}
\centering
\vspace*{-25mm}
\mbox{
\hspace*{-15mm}
\includegraphics[width=0.65\textwidth]{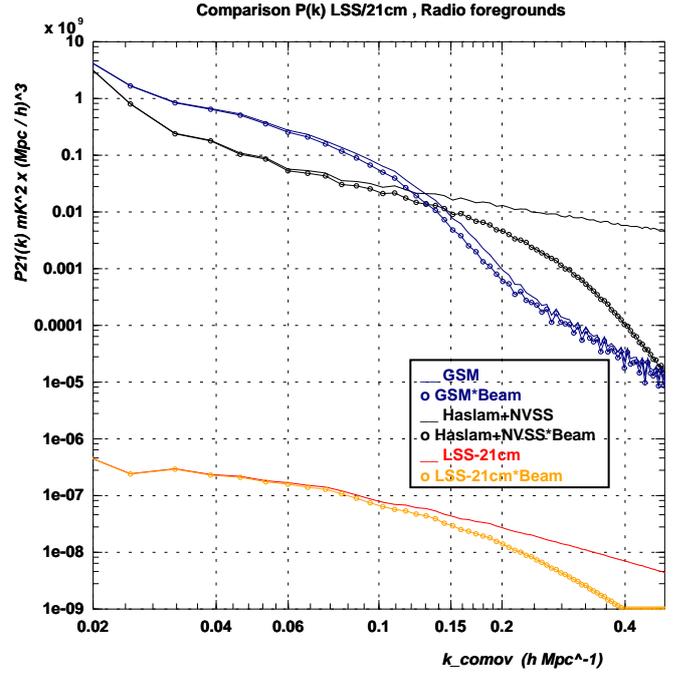}
}
\vspace*{-40mm}
\caption{Comparison of the 21cm LSS power spectrum (red, orange) with the radio foreground power spectrum.
The radio sky power spectrum is shown for the GSM (Model-I) sky model (dark blue), as well as for our simple
model based on Haslam+NVSS (Model-II, black). The curves with circle markers show the power spectrum 
as observed by a perfect instrument with a 25 arcmin (FWHM) gaussian beam.}
\label{pkgsmlss}
\end{figure}

\begin{table}
\centering
\begin{tabular}{|c|c|c|}
\hline
 & mean (K) & std.dev (K) \\
\hline 
Haslam & 2.17 & 0.6 \\
NVSS & 0.13 & 7.73 \\
Haslam+NVSS & 2.3 & 7.75 \\
(Haslam+NVSS)*Lobe(35') & 2.3 & 0.72 \\
GSM & 2.1 & 0.8 \\
\hline
\end{tabular}
\caption{ Mean temperature and standard deviation for the different sky brightness 
data cubes computed for this study (see table \ref{skycubechars} for sky cube resolution and size).}
\label{sigtsky}
\end{table}

we have computed the power spectrum for the 21cm-LSS sky temperature cube, as well 
as for the radio foreground temperature cubes obtained from the two 
models. We have also computed the power spectrum on sky brightness temperature
cubes, as measured by a perfect instrument having a 25 arcmin (FWHM) gaussian beam.
The resulting computed power spectra are shown on figure \ref{pkgsmlss}. 
The GSM model has more large scale power compared to our simple Haslam+NVSS model, 
while it lacks power at higher spatial frequencies. The mode mixing due to 
frequency dependent response will thus be stronger in Model-II (Haslam+NVSS) 
case. It can also be seen that the radio foreground power spectrum is more than
$\sim 10^6$ times higher than the 21 cm signal from large scale structures. This corresponds 
to the factor $\sim 10^3$ of the sky brightness temperature fluctuations ($\sim$ K),
compared to the mK LSS signal.  

It should also be noted that in section 3, we presented the different instrument 
configuration noise levels after {\em correcting or deconvolving} the instrument response. The LSS 
power spectrum is recovered unaffected in this case, while the noise power spectrum 
increases at high k values (small scales). In practice, clean deconvolution is difficult to 
implement for real data and the power spectra presented in this section are NOT corrected 
for the instrumental response.  The observed structures have thus a scale dependent damping
according to the instrument response, while the instrument noise is flat (white noise or scale independent). 

\subsection{ Instrument response and LSS signal extraction }
\label{recsec}
The {\it observed} data cube is obtained from the sky brightness temperature 3D map 
$T_{sky}(\alpha, \delta, \nu)$ by applying the frequency or wavelength dependent instrument response
${\cal R}(u,v,\lambda)$. 
We have considered the simple case where  the instrument response is constant throughout the survey area, or independent 
of the sky direction.
For each frequency $\nu_k$ or wavelength $\lambda_k=c/\nu_k$ : 
\begin{enumerate}
\item Apply a 2D Fourier transform to compute sky angular Fourier amplitudes
$$ T_{sky}(\alpha, \delta, \lambda_k) \rightarrow \mathrm{2D-FFT} \rightarrow {\cal T}_{sky}(u, v, \lambda_k)$$ 
\item Apply instrument response in the angular wave mode plane. We use here the normalized instrument response
$ {\cal R}(u,v,\lambda_k)  \lesssim 1$. 
$$  {\cal T}_{sky}(u, v, \lambda_k)  \longrightarrow {\cal T}_{sky}(u, v, \lambda_k) \times {\cal R}(u,v,\lambda_k) $$
\item Apply inverse 2D Fourier transform to compute the measured sky brightness temperature map, 
without instrumental (electronic/$\Tsys$) white noise:
$$ {\cal T}_{sky}(u, v, \lambda_k) \times {\cal R}(u,v,\lambda)   
\rightarrow \mathrm{Inv-2D-FFT} \rightarrow T_{mes1}(\alpha, \delta, \lambda_k) $$ 
\item Add white noise (gaussian fluctuations) to the pixel map temperatures to obtain 
the measured sky brightness temperature $T_{mes}(\alpha, \delta, \nu_k)$. 
We have also considered that the system temperature and thus the 
additive white noise level was independent of the frequency or wavelength.   
\end{enumerate} 
The LSS signal extraction depends indeed on the white noise level. 
The results shown here correspond to the (a) instrument configuration, a packed array of 
$11 \times 11 = 121$ dishes  (5 meter diameter), with a white noise level corresponding 
to $\sigma_{noise} = 0.25 \mathrm{mK}$ per $3 \times 3 \mathrm{arcmin^2} \times 500$ kHz 
cell.

A brief description of the simple component separation procedure that we have applied is given here:
\begin{enumerate}
\item The measured sky brightness temperature is first {\em corrected} for the frequency dependent 
beam effects through a convolution by a fiducial frequency independent beam. This {\em correction}
corresponds to a smearing or degradation of the angular resolution. We assume 
that we have a perfect knowledge of the intrinsic instrument response, up to a threshold numerical level 
of about $ \gtrsim 1 \%$ for  ${\cal R}(u,v,\lambda)$. We recall that this is the normalized instrument response,
${\cal R}(u,v,\lambda) \lesssim 1$. 
$$  T_{mes}(\alpha, \delta, \nu) \longrightarrow T_{mes}^{bcor}(\alpha,\delta,\nu) $$ 
The virtual target instrument has a beam width larger than the worst real instrument beam,
i.e at the lowest observed frequency.  
 \item For each sky direction $(\alpha, \delta)$, a power law $T = T_0 \left( \frac{\nu}{\nu_0} \right)^b$
 is fitted to the beam-corrected brightness temperature. The fit is done through a linear $\chi^2$ fit in 
the $\lgd ( T ) , \lgd (\nu)$ plane and we show here the results for a pure power law (P1) 
or modified power law (P2):
\begin{eqnarray*} 
P1 & :  & \lgd ( T_{mes}^{bcor}(\nu) ) = a + b \, \lgd ( \nu / \nu_0 ) \\
P2 & :  & \lgd ( T_{mes}^{bcor}(\nu) ) = a + b \, \lgd ( \nu / \nu_0 ) + c \, \lgd ( \nu/\nu_0 ) ^2 
\end{eqnarray*}
where $b$ is the power law index and  $T_0 = 10^a$ is the brightness temperature at the 
reference frequency $\nu_0$:
\item The difference between the beam-corrected sky temperature and the fitted power law
$(T_0(\alpha, \delta), b(\alpha, \delta))$ is our extracted 21 cm LSS signal.
\end{enumerate}

\begin{figure*}
\centering
\vspace*{-25mm}
\mbox{
\hspace*{-20mm}
\includegraphics[width=1.15\textwidth]{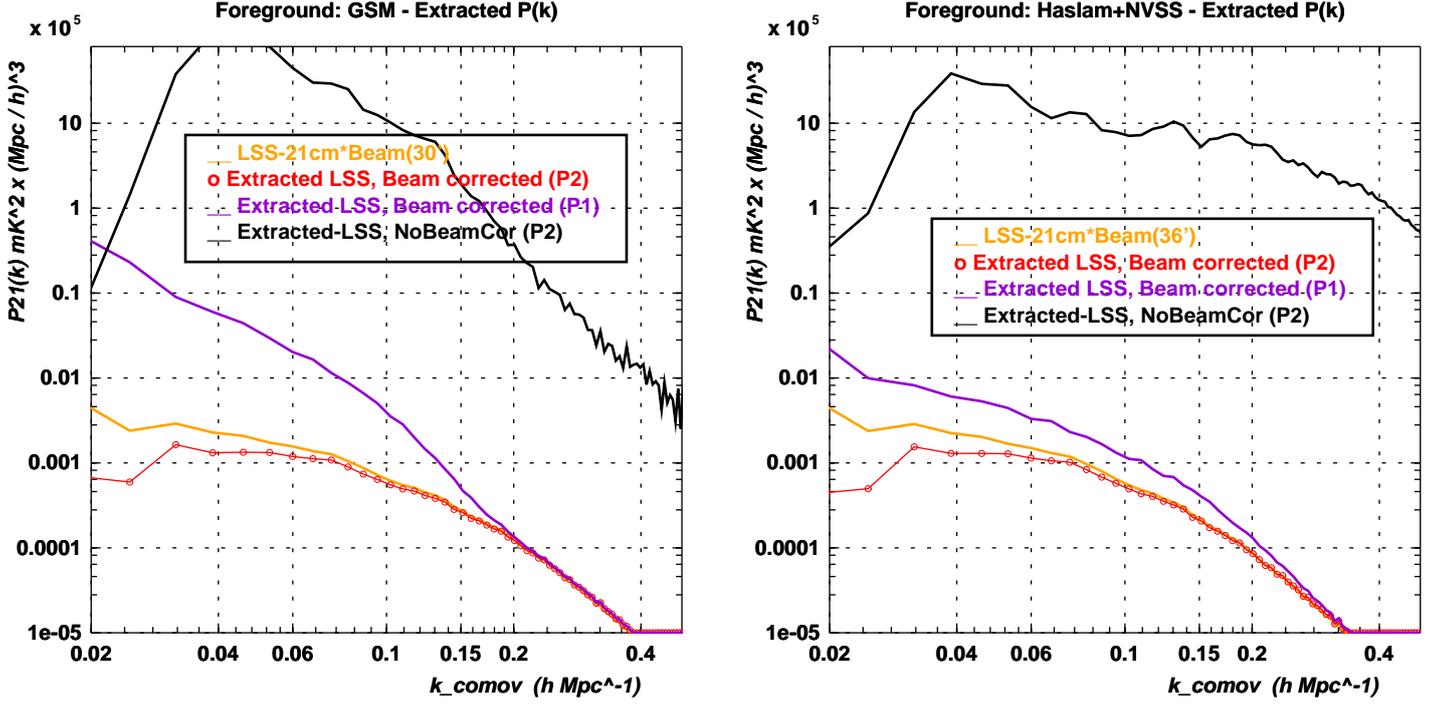}
}
\vspace*{-35mm}
\caption{Recovered power spectrum of the 21cm LSS temperature fluctuations, separated from the 
continuum radio emissions at $z \sim 0.6$, for the instrument configuration (a), $11\times11$ 
packed array interferometer.
Left: GSM/Model-I , right: Haslam+NVSS/Model-II. black curve shows the residual after foreground subtraction,
corresponding to the 21 cm signal, WITHOUT applying the beam correction. Red curve shows the recovered 21 cm 
signal power spectrum, for P2 type fit of the frequency dependence of the radio continuum, and violet curve is the P1 fit (see text). The orange/yellow curve shows the original 21 cm signal power spectrum, smoothed with a perfect, frequency independent gaussian beam. }
\label{extlsspk}
\end{figure*}

Figure \ref{extlsspk} shows the performance of this procedure at a redshift $\sim 0.6$, 
for the two radio sky models used here: GSM/Model-I and Haslam+NVSS/Model-II. The 
21 cm LSS power spectrum, as seen by a perfect instrument with a 25 arcmin (FWHM) 
gaussian frequency independent beam is shown in orange (solid line), 
and the extracted power spectrum, after beam {\em correction} 
and foreground separation with second order polynomial fit (P2) is shown in red (circle markers).
We have also represented the obtained power spectrum without applying the beam correction (step 1 above),
or with the first order polynomial fit (P1). 

\begin{figure*}
\centering
\vspace*{-20mm}
\mbox{
\hspace*{-25mm}
\includegraphics[width=1.20\textwidth]{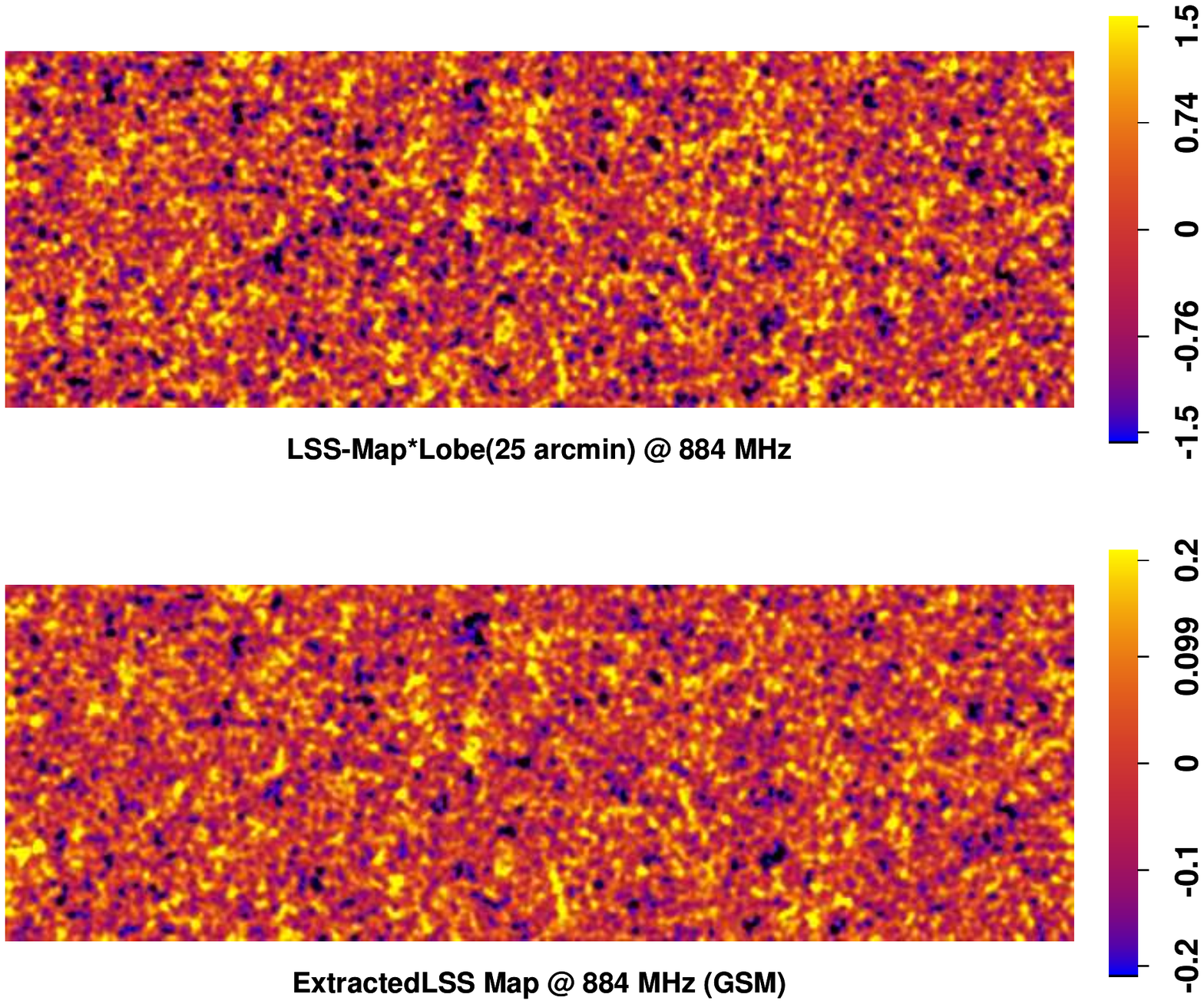}
}
\vspace*{-25mm}
\caption{Comparison of the original 21 cm LSS temperature map @ 884 MHz ($z \sim 0.6$), smoothed
with 25 arc.min (FWHM) beam (top), and the recovered LSS map, after foreground subtraction for Model-I (GSM) (bottom),  for the instrument configuration (a), $11\times11$ packed array interferometer.
Notice the difference between the temperature color scales (mK)  for the top and bottom maps. }
\label{extlssmap}
\end{figure*}

Figure \ref{extlssmap} shows a comparison of  the original 21 cm brightness temperature map at 884 MHz 
with the recovered 21 cm map, after subtraction of the radio continuum component. It can be seen that structures 
present in the original map have been correctly recovered, although the amplitude of the temperature 
fluctuations on the recovered map is significantly smaller (factor $\sim 5$) than in the original map. This is mostly 
due to the damping of the large scale ($k \lesssim 0.04 h \mathrm{Mpc^{-1}} $) due the poor interferometer 
response at large angle    ($\theta \gtrsim 4^\circ $). 

We have shown that it should be possible to measure the red shifted 21 cm emission fluctuations in the 
presence of the strong radio continuum signal, provided that this latter has a smooth frequency dependence.
However, a rather  precise knowledge of the instrument beam and the beam {\em correction} 
or smearing procedure described here  are key ingredient for recovering the 21 cm LSS power spectrum. 
It is also important to note that while it is enough to correct the beam to the lowest resolution instrument beam 
($\sim 30'$ or $D \sim 50$ meter @ 820 MHz) for the GSM sky model, a stronger beam correction 
has to be applied (($\sim 36'$ or $D \sim 40$ meter @ 820 MHz) for the Model-II to reduce 
significantly the ripples from bright radio sources. 
We have also applied the same procedure to simulate observations and LSS signal extraction for an instrument 
with a frequency dependent gaussian beam shape. The mode mixing effect is greatly reduced for
such a smooth beam, compared to the more complex instrument response 
${\cal R}(u,v,\lambda)$ used for the results shown in figure \ref{extlsspk}.

\subsection{$P_{21}(k)$ measurement transfer function} 
\label{tfpkdef}
The recovered red shifted 21 cm emission power spectrum $P_{21}^{rec}(k)$ suffers a number of distortions, mostly damping,
 compared to the original $P_{21}(k)$ due to  the instrument response and the component separation procedure.
We expect damping at small scales, or larges $k$, due to the finite instrument size, but also at large scales, small $k$,
if total power measurements (auto-correlations) are not used in the case of interferometers. 
The sky reconstruction and the component separation introduce additional filtering and distortions.
Ideally, one has to define a power spectrum measurement response or {\it transfer function} in the 
radial direction,  ($\lambda$ or redshift, $\TrF(k_\parallel)$) and in the transverse plane ( $\TrF(k_\perp)$ ). 
The real transverse plane transfer function might even be anisotropic. 

However, in the scope of the present study, we define an overall transfer function $\TrF(k)$ as the ratio of the 
recovered 3D power spectrum $P_{21}^{rec}(k)$ to the original $P_{21}(k)$:
\begin{equation}
\TrF(k) = P_{21}^{rec}(k) / P_{21}(k) 
\end{equation}

\begin{figure*}
\centering
\vspace*{-30mm}
\mbox{
\hspace*{-20mm}
\includegraphics[width=1.15\textwidth]{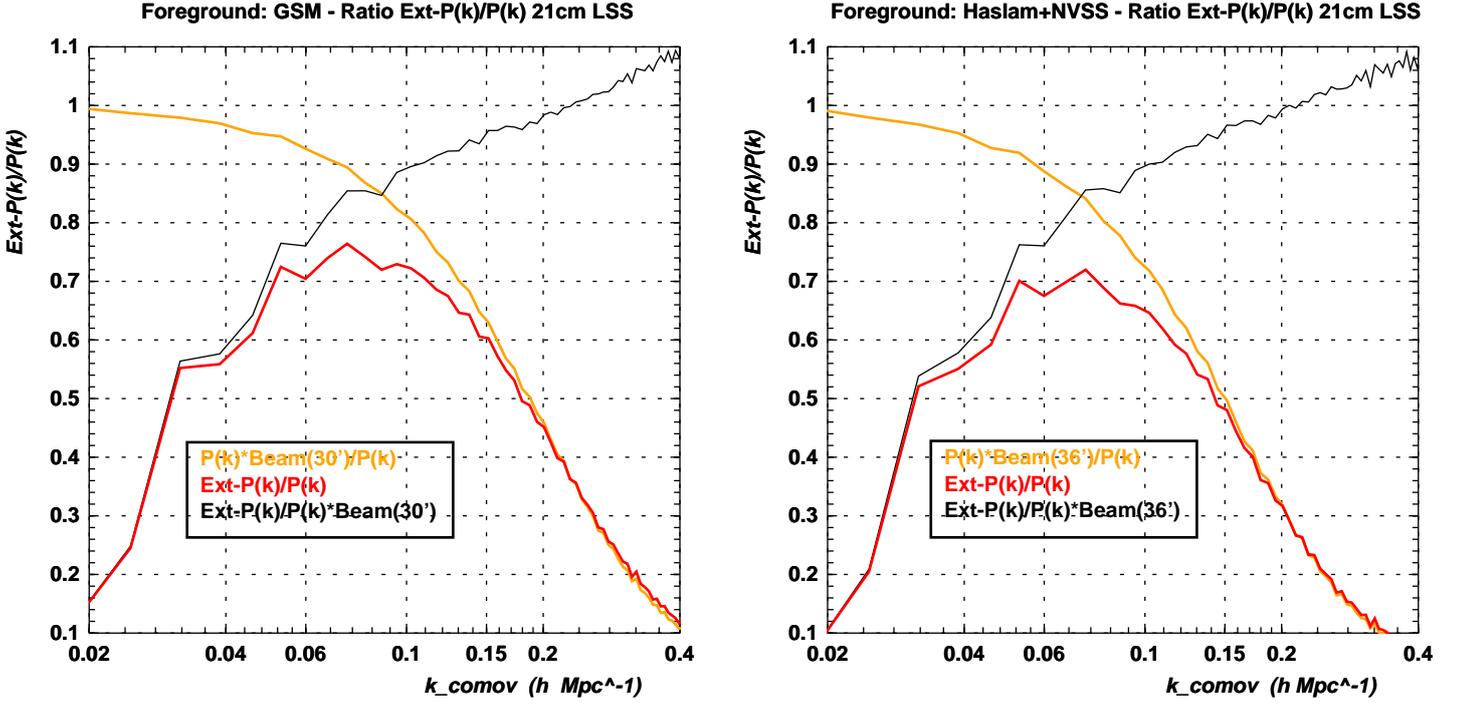}
}
\vspace*{-35mm}
\caption{Ratio of the reconstructed or extracted 21cm power spectrum, after foreground removal, to the initial 21 cm power spectrum, $\TrF(k) = P_{21}^{rec}(k) / P_{21}(k) $, at $z \sim 0.6$,  for the instrument configuration (a), $11\times11$ packed array interferometer.
Left: GSM/Model-I , right: Haslam+NVSS/Model-II.  }
\label{extlssratio}
\end{figure*}

Figure \ref{extlssratio} shows this overall transfer function for the simulations and component 
separation performed here, around $z \sim 0.6$, for the instrumental setup (a), a filled array of 121 $D_{dish}=5$ m dishes.
The orange/yellow curve shows the ratio $P_{21}^{smoothed}(k)/P_{21}(k)$ of the computed to the original 
power spectrum, if the original LSS temperature cube is smoothed by the frequency independent target beam 
FWHM=30' for the GSM simulations (left), 36' for Model-II (right). This orange/yellow 
curve shows the damping effect due to the finite instrument size at small scales ($k \gtrsim 0.1 \, h \, \mathrm{Mpc^{-1}}, \theta \lesssim 1^\circ$).  
The recovered power spectrum suffers also significant damping at large scales $k \lesssim 0.05 \, h \, \mathrm{Mpc^{-1}}, $ due to poor interferometer 
response at large angles ($ \theta \gtrsim 4^\circ-5^\circ$), as well as to the filtering of radial or longitudinal Fourier modes along 
the frequency or redshift direction ($k_\parallel$) by the component separation algorithm. 
The red curve shows the ratio of $P(k)$ computed on the recovered or extracted 21 cm LSS signal, to the original 
LSS temperature cube $P_{21}^{rec}(k)/P_{21}(k)$ and corresponds to the transfer function $\TrF(k)$ defined above, 
for $z=0.6$ and instrument setup (a). 
The black (thin line) curve shows the ratio of recovered to the smoothed 
power spectrum $P_{21}^{rec}(k)/P_{21}^{smoothed}(k)$. This latter ratio (black curve) exceeds one for $k \gtrsim 0.2$, which is 
due to the noise or system temperature. It should be stressed that the simulations presented in this section were 
focused on the study of the radio foreground effects and have been carried intently with a very low instrumental noise level of 
$0.25$ mK per pixel, corresponding to several years of continuous observations ($\sim 10$ hours per $3' \times 3'$ pixel).

This transfer function is well represented by the analytical form:
\begin{equation}
\TrF(k) = \sqrt{ \frac{ k-k_A}{ k_B}  } \times \exp \left( - \frac{k}{k_C} \right) 
\label{eq:tfanalytique}
\end{equation} 

We have performed  simulation of observations and radio foreground subtraction using 
the procedure described here for different redshifts and instrument configurations, in particular 
for the (e) configuration with 400 five-meter dishes. As the synchrotron and radio source strength 
increases quickly with decreasing frequency, we have seen that recovering the 21 cm LSS signal 
becomes difficult for larger redshifts, in particular for $z \gtrsim 2$. 

\begin{figure}
\centering
\vspace*{-25mm}
\mbox{
\hspace*{-10mm}
\includegraphics[width=0.55\textwidth]{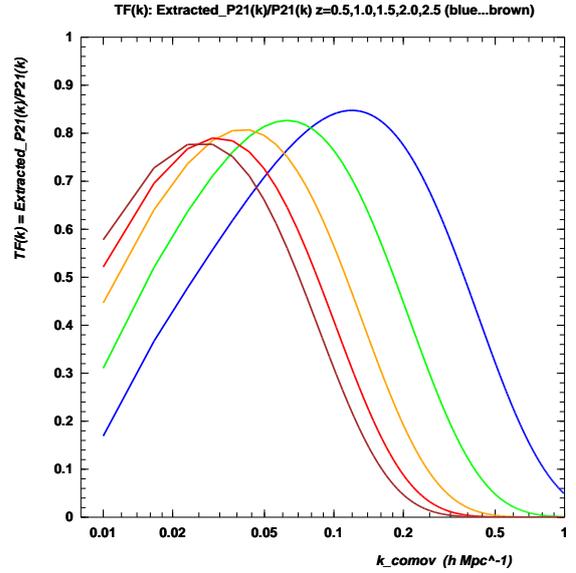}
}
\vspace*{-30mm}
\caption{Fitted/smoothed  transfer function $\TrF(k)$ obtained for the recovered 21 cm power spectrum at different redshifts, 
$z=0.5 , 1.0 , 1.5 , 2.0 , 2.5$ for the instrument configuration (e), $20\times20$ packed array interferometer. }
\label{tfpkz0525}
\end{figure}

We have determined the transfer function parameters of eq. \ref{eq:tfanalytique} $k_A, k_B, k_C$ 
for setup (e) for three redshifts, $z=0.5, 1 , 1.5$, and then extrapolated the value of the parameters 
for redshift $z=2, 2.5$. The value of the parameters are grouped in table \ref{tab:paramtfk} 
and the smoothed transfer functions are shown on  figure \ref{tfpkz0525}. 

\begin{table}[hbt]
\begin{center}
\begin{tabular}{|c|ccccc|}
\hline 
\hspace{2mm} z \hspace{2mm} & \hspace{2mm} 0.5 \hspace{2mm}  & \hspace{2mm} 1.0 \hspace{2mm} &
\hspace{2mm} 1.5 \hspace{2mm} & \hspace{2mm} 2.0 \hspace{2mm}  & \hspace{2mm} 2.5 \hspace{2mm} \\
\hline 
$k_A$ & 0.006 & 0.005 & 0.004 & 0.0035 & 0.003 \\
$k_B$ & 0.038 & 0.019 & 0.012 & 0.0093 & 0.008 \\
$k_C$ & 0.16   & 0.08   & 0.05   & 0.038   & 0.032 \\
\hline 
\end{tabular}
\end{center}
\caption{Value of the parameters for the transfer function (eq. \ref{eq:tfanalytique}) at different redshift 
for instrumental setup (e), $20\times20$ packed array interferometer.  }
\label{tab:paramtfk}
\end{table}

\section{Sensitivity to cosmological parameters}
\label{cosmosec}

The impact of the various telescope configurations on the sensitivity for 21 cm 
power spectrum measurement has been discussed in section \ref{pkmessens}. 
Fig. \ref{figpnoisea2g} shows the noise power spectra, and allows us to rank visually the configurations 
in terms of instrument noise contribution to P(k) measurement. 
The differences in $P_{noise}$ will translate into differing precisions
in the reconstruction of the BAO peak positions and in
the estimation of cosmological parameters. In addition, we have seen (sec. \ref{recsec})
that subtraction of continuum radio emissions, Galactic synchrotron and radio sources, 
has also an effect on the measured 21 cm power spectrum. 
In this paragraph, we present our method and the results for the precisions on the estimation 
of Dark Energy parameters, through a radio survey of the redshifted 21 cm emission of LSS, 
with an instrumental setup similar to the (e) configuration (sec. \ref{instrumnoise}), 400 five-meter diameter 
dishes, arranged into a filled $20 \times 20$ array.

\subsection{BAO peak precision}

In order to estimate the precision with which BAO peak positions can be
measured, we  used a method similar to the one established in 
\citep{blake.03} and \citep{glazebrook.05}.

To this end, we  generated reconstructed power spectra $P^{rec}(k)$ for
 slices of Universe with a quarter-sky coverage and a redshift depth,
 $\Delta z=0.5$ for  $0.25<z<2.75$. 
The peaks in the generated spectra were then determined by a 
fitting procedure and the reconstructed peak positions compared with the 
generated peak positions.
The reconstructed power spectrum used in the simulation is  
the sum of the expected \HI signal term, corresponding to equations \ref{eq:pk21z} and \ref{eq:tbar21z}, 
damped by the transfer function $\TrF(k)$ (Eq. \ref{eq:tfanalytique} , table \ref{tab:paramtfk})
and a white noise component $P_{noise}$ calculated according to the equation \ref{eq:pnoiseNbeam}, 
established in section \ref{instrumnoise} with $N=400$:
\begin{equation}
 P^{rec}(k) = P_{21}(k) \times \TrF(k) + P_{noise} 
\end{equation}
where the different terms ($P_{21}(k) , \TrF(k), P_{noise}$) depend on the slice redshift.  
The expected 21 cm power spectrum $P_{21}(k)$ has been generated according to the formula:
\begin{eqnarray}
\label{eq:signal}
\frac{P_{21}(\kperp,\kpar)}{P_{ref}(\kperp,\kpar)} = 
1\; + 
\hspace*{40mm}
\nonumber
\\ \hspace*{20mm}
A\, k \exp \bigl( -(k/\tau)^\alpha\bigr)
\sin\left( 2\pi\sqrt{\frac{\kperp^2}{\koperp^2} + 
\frac{\kpar^2}{\kopar^2}}\;\right)
\end{eqnarray}
where $k=\sqrt{\kperp^2 + \kpar^2}$, the parameters $A$, $\alpha$ and $\tau$ 
are adjusted to the  formula presented in 
\citep{eisenhu.98}. $P_{ref}(\kperp,\kpar)$ is the 
envelop curve of the HI power spectrum without baryonic oscillations. 
The parameters $\koperp$ and $\kopar$ 
are the inverses of the oscillation periods in k-space. 
The following values have been used for these 
parameters for the results presented here: $A=1.0$, $\tau=0.1 \, \hMpcm$, 
$\alpha=1.4$ and $\koperp=\kopar=0.060 \, \hMpcm$.

Each simulation is performed for a given set of parameters 
which are: the system temperature,$\Tsys$, an observation time, 
$t_{obs}$, an average redshift and a redshift depth, $\Delta z=0.5$. 
Then,  each simulated  power spectrum  is fitted with a two dimensional 
normalized function $P_{tot}(\kperp,\kpar)/P_{ref}(\kperp,\kpar)$ which is 
the sum of the signal power spectrum damped by the transfer function and the 
noise power spectrum  multiplied by a 
linear term,  $a_0+a_1k$. The upper limit $k_{max}$ in $k$ of the fit 
corresponds to the approximate position of the linear/non-linear transition. 
This limit is established on the basis of the criterion discussed in  
\citep{blake.03}. 
In practice, we used for the redshifts 
$z=0.5,\,\, 1.0$ and  $1.5$ respectively $k_{max}= 0.145 \hMpcm,\,\, 0.18\hMpcm$ 
and $0.23 \hMpcm$. 
 
\begin{figure}[htbp]
\begin{center}
\includegraphics[width=8.5cm]{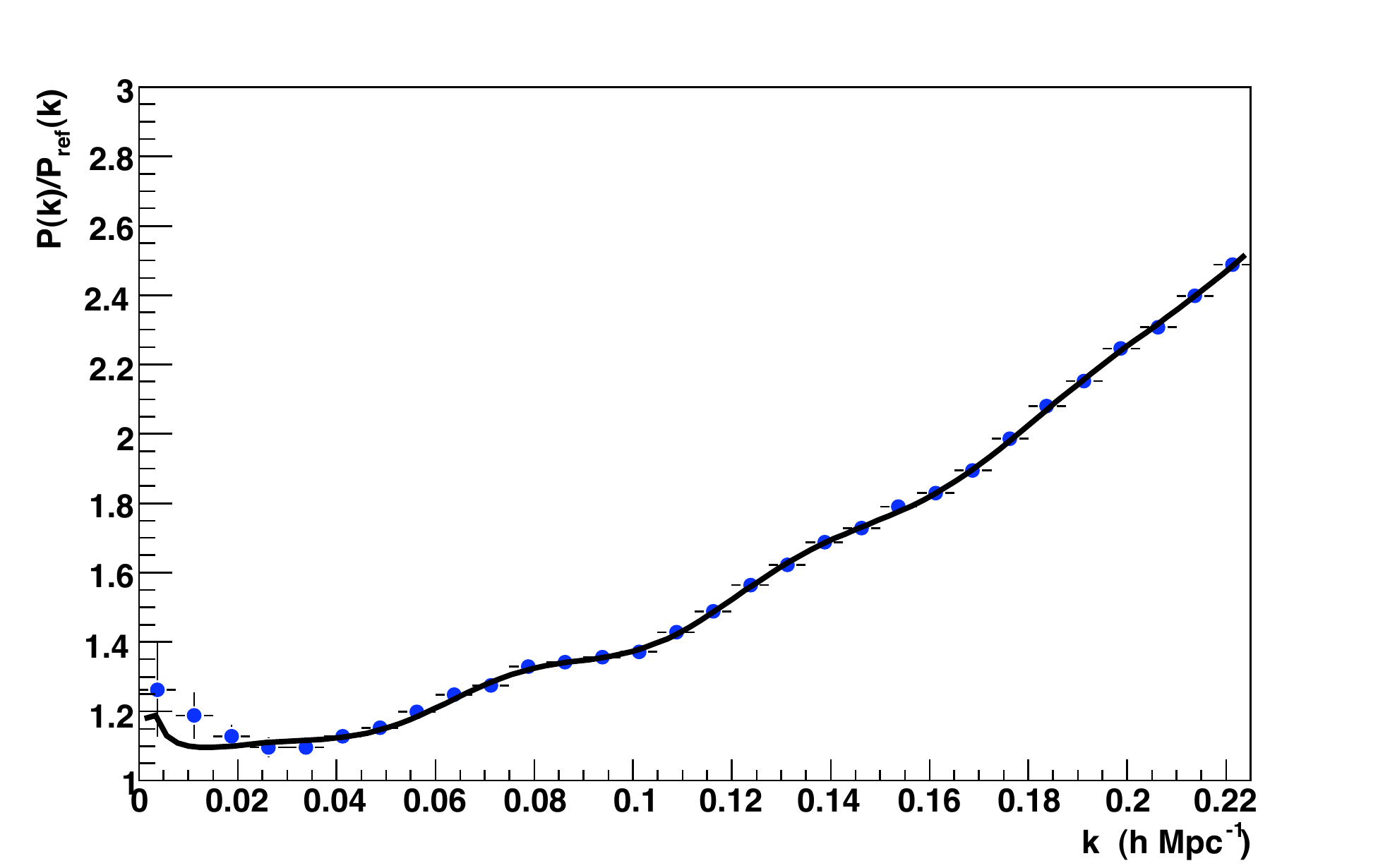}
\caption{1D projection of the power spectrum for one simulation. 
The \HI power spectrum is divided by an envelop curve $P(k)_{ref}$ 
corresponding to the power spectrum without baryonic oscillations. 
The dots represents one simulation for a "packed" array of cylinders  
with a system temperature,$T_{sys}=50$K, an observation time, 
$T_{obs}=$ 1 year, 
a solid angle of $1\pi sr$,
an average redshift, $z=1.5$ and a redshift depth, $\Delta z=0.5$. 
The solid line is the result of the fit to the data.}
\label{fig:fitOscill}
\end{center}
\end{figure}

\begin{figure}[htbp]
\begin{center}
\includegraphics[width=9.0cm]{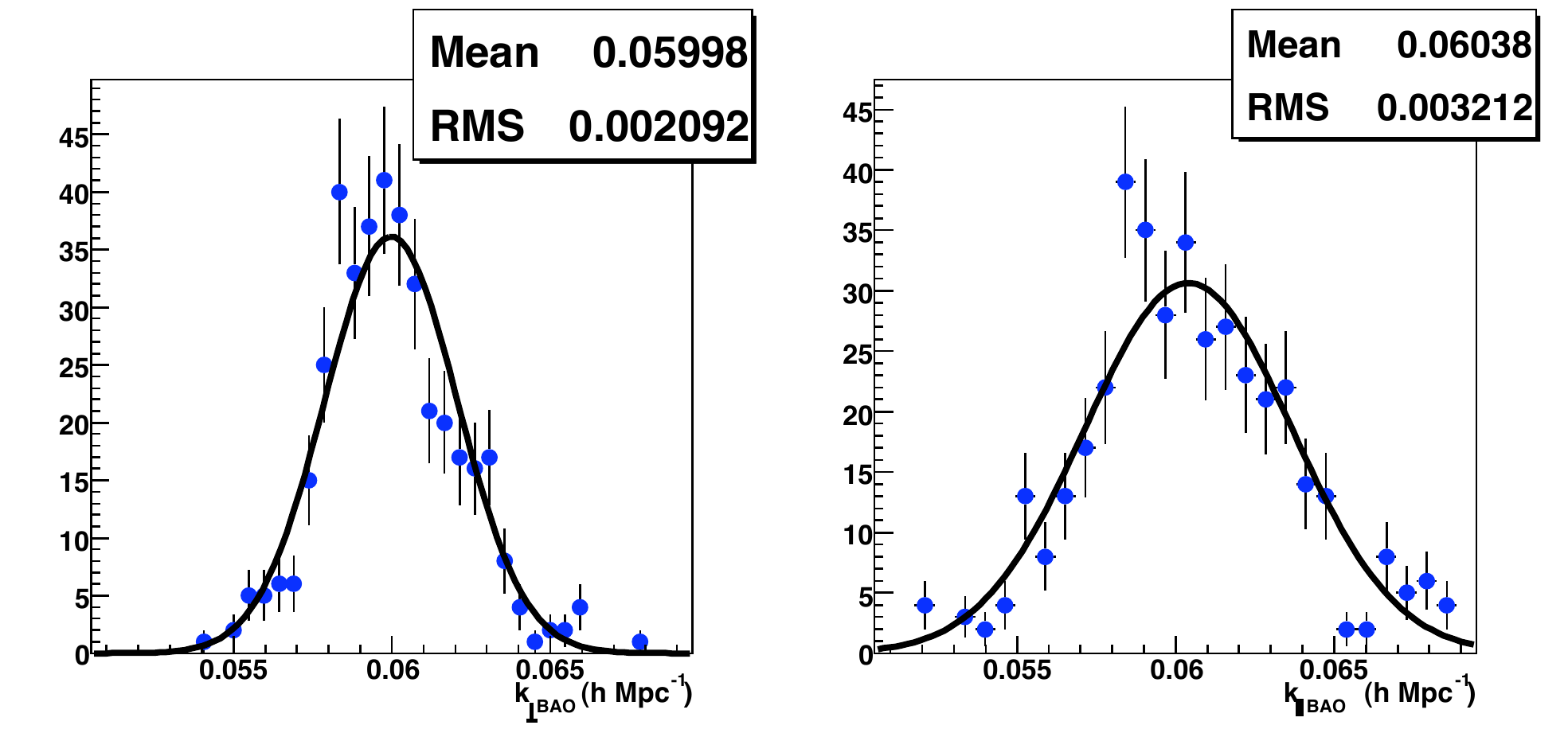}
\caption{ Distributions of the reconstructed 
wavelength  $\koperp$ and $\kopar$ 
respectively, perpendicular and parallel to the line of sight
for simulations as in Fig. \ref{fig:fitOscill}. 
The fit by a Gaussian of the distribution (solid line) gives the 
width of the distribution  which represents the statistical error 
expected on these parameters.}
\label{fig:McV2}
\end{center}
\end{figure}

Figure \ref{fig:fitOscill} shows the result of the fit for 
one of these simulations. 
Figure \ref{fig:McV2} histograms the recovered values of  $\koperp$ and $\kopar$
for 100 simulations.
The widths of the two distributions give an estimate 
of the statistical errors.

\begin{figure}[htbp]
\begin{center}
\includegraphics[width=8.5cm]{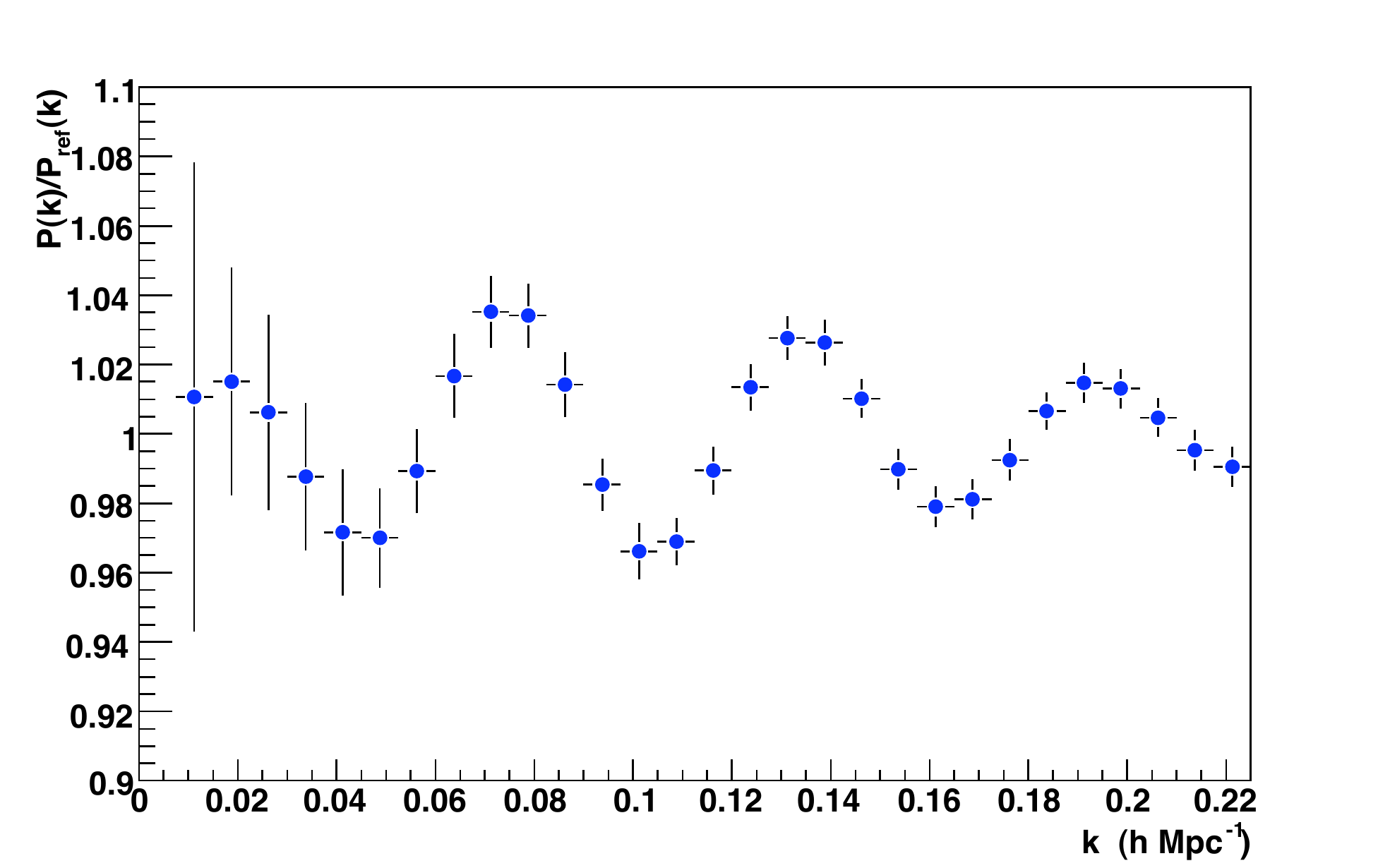}
\caption{1D projection of the power spectrum averaged over 100 simulations
of the packed cylinder array $b$. 
The simulations are performed for the following conditions: a system 
temperature, $T_{sys}=50$K, an observation time, $T_{obs}=1$ year, 
a solid angle of $1 \pi sr$,
an average redshift, $z=1.5$ and a redshift depth, $\Delta z=0.5$. 
The \HI power spectrum is divided by an envelop curve $P(k)_{ref}$ 
corresponding to the power spectrum without baryonic oscillations 
and the background estimated by a fit is subtracted. The errors are 
the RMS of the 100 distributions for each $k$ bin and the dots are 
the mean of the distribution for each $k$ bin. }
\label{fig:AverPk}
\end{center}
\end{figure}

In addition, in the fitting procedure, both the parameters modeling the 
signal $A$, $\tau$, $\alpha$ and the parameter correcting the noise power 
spectrum $(a_0,a_1)$ are floated to take into account the possible 
ignorance  of the signal shape and the uncertainties in the 
computation of the noise power spectrum. 
In this way, we can correct possible imperfections and the 
systematic uncertainties are directly propagated to statistical errors 
on the relevant parameters  $\koperp$ and $\kopar$. By subtracting the 
fitted noise contribution to each simulation, the baryonic oscillations 
are clearly observed, for instance, on Fig.~\ref{fig:AverPk}.

In our comparison of the various configurations, we have considered 
the following cases for $\Delta z=0.5$ slices with $0.25<z<2.75$.
\begin{itemize}
\item {\it Simulation without electronics noise}: the statistical errors on the power 
spectrum are directly related to the number of modes in the surveyed volume $V$ corresponding to 
 $\Delta z=0.5$ slice with the solid angle $\Omega_{tot}$ = 1 $\pi$ sr. 
The number of mode $N_{\delta k}$ in the wave number interval $\delta k$ can be written as:
\begin{equation}
V  =  \frac{c}{H(z)} \Delta z  \times (1+z)^2 \dang^2  \Omega_{tot} \hspace{10mm}
N_{\delta k}  =  \frac{ V }{4 \pi^2} k^2 \delta k 
\end{equation}   
\item {\it Noise}: we add the instrument noise as a constant term $P_{noise}$ as described in Eq. 
\ref {eq:pnoiseNbeam}. Table \ref{tab:pnoiselevel} gives the white noise level for 
$\Tsys = 50 \mathrm{K}$ and one year total observation time to survey $\Omega_{tot}$ = 1 $\pi$ sr.
\item {\it Noise with transfer function}: we take into account of the interferometer and radio foreground
subtraction represented as the measured P(k) transfer function $T(k)$ (section \ref{tfpkdef}), as 
well as instrument noise $P_{noise}$.
\end{itemize}

\begin{table}
\begin{tabular}{|l|ccccc|}
\hline
z & \hspace{1mm} 0.5 \hspace{1mm} &  \hspace{1mm} 1.0 \hspace{1mm} &
\hspace{1mm} 1.5 \hspace{1mm} &  \hspace{1mm} 2.0 \hspace{1mm} & \hspace{1mm} 2.5 \hspace{1mm} \\
\hline 
$P_{noise} \, \mathrm{mK^2 \, (Mpc/h)^3}$ &  8.5 & 35 & 75 & 120 & 170 \\
\hline 
\end{tabular}
\caption{Instrument or electronic noise spectral power $P_{noise}$ for a $N=400$ dish interferometer with $\Tsys=50$ K and $t_{obs} =$ 1 year to survey $\Omega_{tot} = \pi$ sr }
\label{tab:pnoiselevel} 
\end{table}

Table \ref{tab:ErrorOnK} summarizes the result. The errors both on $\koperp$ and $\kopar$
decrease as a function of redshift for simulations without electronic noise because the volume of the universe probed is larger. Once we apply the electronics noise, each slice in redshift give comparable results.  Finally, after applying the full reconstruction of the interferometer, the best accuracy is obtained for the first slices in redshift around 0.5 and 1.0 for an identical time of observation. We can optimize the survey by using a different observation time for each slice in redshift. Finally, for a 3 year survey we can split in five observation periods with durations which are 3 months, 3 months, 6 months, 1 year and 1 year respectively for redshift 0.5, 1.0, 1.5, 2.0 and 2.5.

\begin{table*}[ht]
\begin{center}
\begin{tabular}{lc|c c c c c }
\multicolumn{2}{c|}{$\mathbf z$ }& \bf 0.5 & \bf 1.0 &  \bf 1.5 & \bf 2.0 & \bf 2.5 \\
\hline\hline
\bf No Noise & $\sigma(\koperp)/\koperp$  (\%) & 1.8 & 0.8 & 0.6 & 0.5 &0.5\\
 & $\sigma(\kopar)/\kopar$  (\%) & 3.0 & 1.3 & 0.9 &  0.8 & 0.8\\
 \hline
 \bf  Noise without Transfer Function   & $\sigma(\koperp)/\koperp$  (\%) & 2.3 & 1.8 & 2.2 & 2.4 & 2.8\\
 (3-months/redshift)& $\sigma(\kopar)/\kopar$  (\%) & 4.1 & 3.1  & 3.6 & 4.3 & 4.4\\
 \hline
 \bf   Noise with Transfer Function  & $\sigma(\koperp)/\koperp$  (\%) & 3.0 & 2.5 & 3.5 & 5.2 & 6.5 \\
 (3-months/redshift)& $\sigma(\kopar)/\kopar$  (\%) & 4.8 & 4.0 & 6.2 & 9.3 & 10.3\\
 \hline
 \bf  Optimized survey & $\sigma(\koperp)/\koperp$  (\%)   & 3.0 & 2.5 & 2.3 &  2.0 &  2.7\\
 (Observation time :  3 years)& $\sigma(\kopar)/\kopar$  (\%) & 4.8 & 4.0 & 4.1 &  3.6  & 4.3 \\
 \hline
\end{tabular}
\end{center}
\caption{Sensitivity on the measurement of $\koperp$ and $\kopar$ as a 
function of the redshift $z$ for various simulation configuration. 
$1^{\rm st}$ row: simulations without noise with pure cosmic variance; 
$2^{\rm nd}$ 
row: simulations with electronics noise for a telescope with dishes; 
$3^{\rm th}$ row: simulations 
with same electronics noise and with correction with the transfer function ; 
$4^{\rm th}$ row: optimized survey with a total observation time of 3 years (3 months, 3 months, 6 months, 1 year and 1 year respectively for redshift 0.5, 1.0, 1.5, 2.0 and 2.5 ).}
\label{tab:ErrorOnK}
\end{table*}%

\subsection{Expected sensitivity on $w_0$  and $w_a$}

\begin{figure}
\begin{center}
\includegraphics[width=8.5cm]{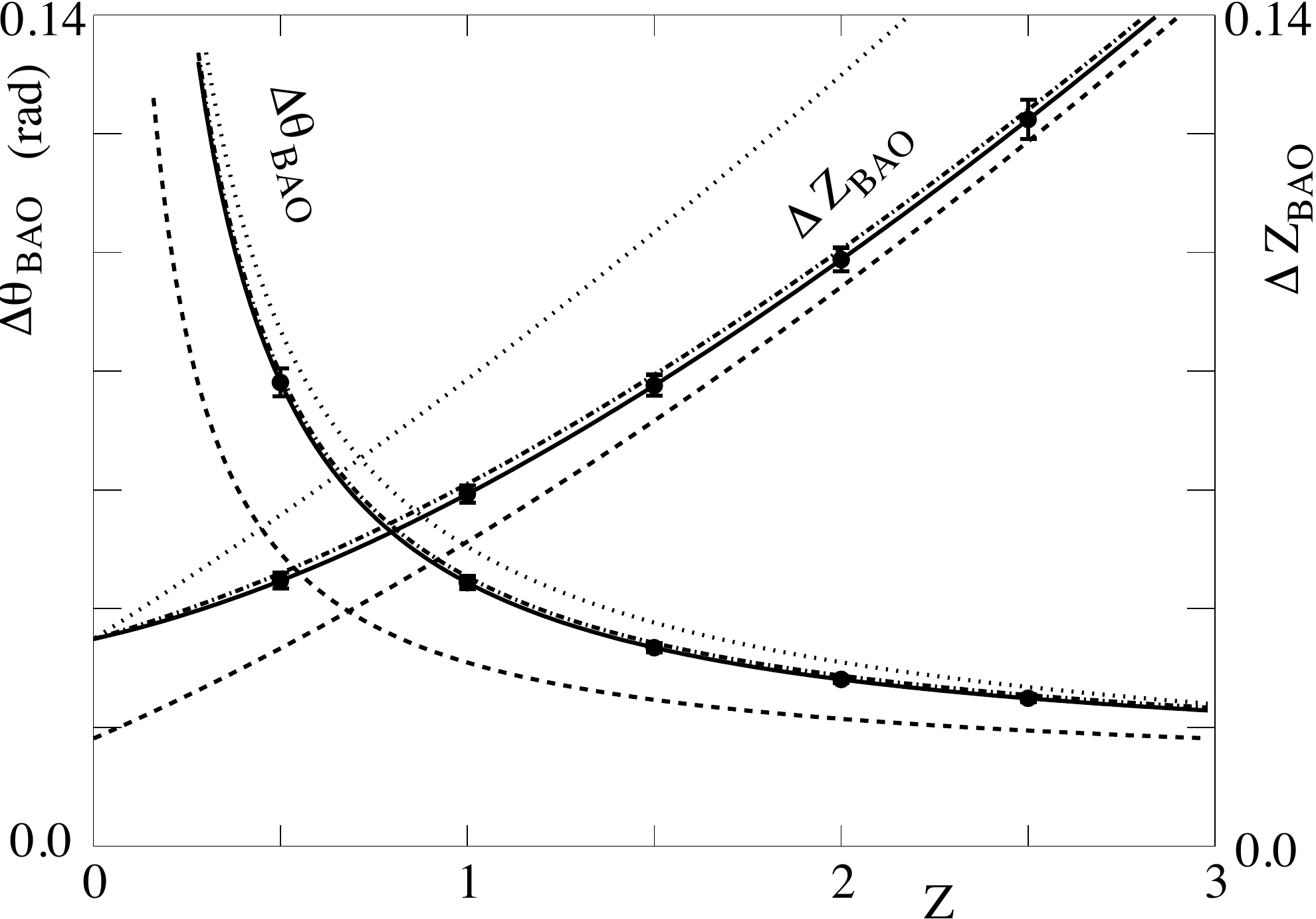}
\caption{
The two ``Hubble diagrams'' for BAO experiments.
The four falling curves give the angular size of the acoustic horizon
(left scale) and the four 
rising curves give the redshift interval of the acoustic horizon (right scale).
The solid lines are for 
$(\Omega_M,\Omega_\Lambda,w)=(0.27,0.73,-1)$,
the dashed  for 
$(1,0,-1)$
the dotted for 
$(0.27,0,-1)$, and
the dash-dotted  for 
$(0.27,0.73,-0.9)$,
The error bars on the solid curve correspond to the four-month run
(packed array)
of Table \ref{tab:ErrorOnK}.
 }
\label{fig:hubble}
\end{center}
\end{figure}

The observations give the \HI power spectrum in
angle-angle-redshift space rather than in real space.
The inverse of the peak positions  in the observed power spectrum therefore  
gives the angular and redshift intervals corresponding to the
sonic horizon.
The peaks in the angular spectrum are proportional to 
$d_T(z)/a_s$ and those in the redshift spectrum to $d_H(z)/a_s$.
$a_s \sim 105  h^{-1} \mathrm{Mpc}$ is the acoustic horizon comoving size at recombination, 
$d_T(z) = (1+z) \dang$ is the comoving angular distance and $d_H=c/H(z)$ is the Hubble distance
(see Eq. \ref{eq:expHz}):
\begin{equation}
d_H = \frac{c}{H(z)} =  \frac{c/H_0}{\sqrt{\Omega_\Lambda+\Omega_m (1+z)^3} }   \hspace{5mm}
d_T = \int_0^z d_H(z) dz 
\label{eq:dTdH}
\end{equation}
The quantities $d_T$, $d_H$ and $a_s$ all depend on
the cosmological parameters.
Figure \ref{fig:hubble} gives the angular and redshift intervals
as a function of redshift for four cosmological models.
The error bars on the lines for
$(\Omega_M,\Omega_\Lambda)=(0.27,0.73)$
correspond to the expected errors 
on the peak positions
taken from Table \ref{tab:ErrorOnK}
for the four-month runs with the packed array.
We see that with these uncertainties, the data would be able to 
measure $w$ at better than the 10\% level.

To estimate the sensitivity 
to parameters describing dark energy equation of 
state, we follow the procedure explained in 
\citep{blake.03}. We can introduce the equation of 
state of dark energy, $w(z)=w_0 + w_a\cdot z/(1+z)$ by 
replacing $\Omega_\Lambda$ in the definition of $d_T (z)$ and $d_H (z)$,
(Eq. \ref{eq:dTdH}) by:
\begin{equation}
\Omega_\Lambda \rightarrow \Omega_{\Lambda} \exp \left[ 3  \int_0^z   
\frac{1+w(z^\prime)}{1+z^\prime } dz^\prime  \right]
\end{equation}
where $\Omega_{\Lambda}^0$ is the present-day dark energy fraction with 
respect to the critical density. 
Using the relative errors on  $\koperp$ and  $\kopar$ given in 
Tab.~\ref{tab:ErrorOnK}, we can compute the Fisher matrix for  
five cosmological parameter: $(\Omega_m, \Omega_b, h, w_0, w_a)$. 
Then, the combination of this BAO Fisher 
matrix with the Fisher matrix obtained for Planck mission, allows us to 
compute the errors on dark energy parameters.
The Planck Fisher matrix is
obtained for the 8 parameters (assuming a flat universe): 
$\Omega_m$, $\Omega_b$, $h$, $w_0$, $w_a$,
$\sigma_8$, $n_s$ (spectral index of the primordial power spectrum) and
$\tau$  (optical depth to the last-scatter surface).

\begin{figure}[htbp]
\begin{center}
\includegraphics[width=0.55\textwidth]{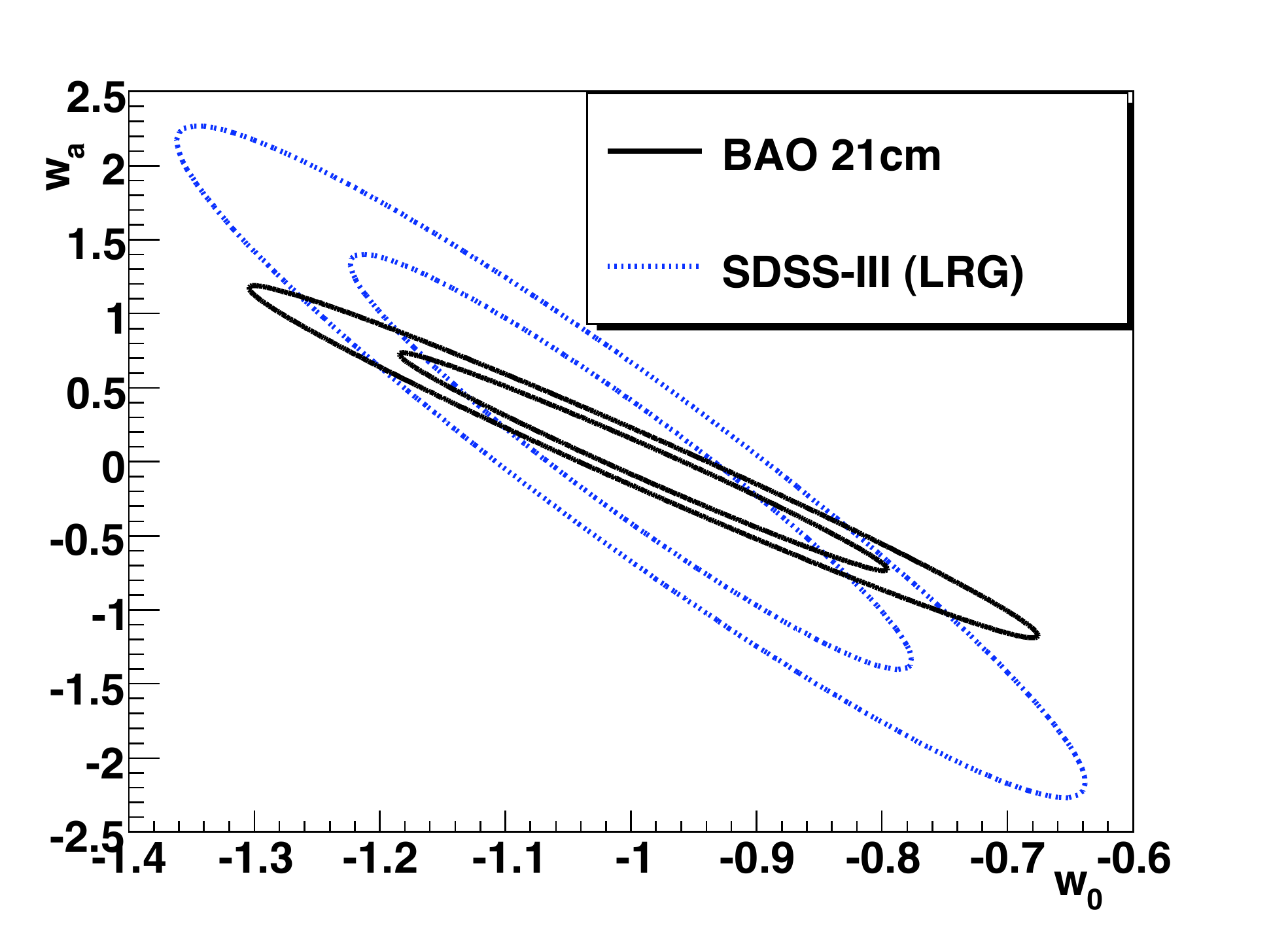}
\caption{$1\sigma$ and $2\sigma$ confidence level contours  in the 
parameter plane $(w_0,w_a)$ for two BAO projects:   SDSS-III (LRG) project 
(blue dotted line), 21 cm project with HI intensity mapping (black solid line).}
\label{fig:Compw0wa}
\end{center}
\end{figure}

For an optimized project over a redshift range, $0.25<z<2.75$, with a total 
observation time of 3 years, the packed 400-dish interferometer array has a 
precision of  12\% on $w_0$ and 48\% on $w_a$. 
The  Figure of Merit, the inverse of the area in the 95\% confidence level 
contours  is 38.
 Finally, Fig.~\ref{fig:Compw0wa} 
shows a comparison of different BAO projects, with a set of priors on 
$(\Omega_m, \Omega_b, h)$ corresponding to the expected precision on 
these parameters in early 2010's. This BAO project based on \HI intensity 
mapping is clearly competitive with the current generation of optical 
surveys such as SDSS-III \citep{sdss3}.

\section{Conclusions}
The 3D mapping of redshifted 21 cm emission though {\it Intensity Mapping} is a novel and complementary 
approach to optical surveys to study the statistical properties of the large scale structures in the universe
up to redshifts $z \lesssim 3$. A radio instrument with large instantaneous field of view 
(10-100 deg$^2$) and large bandwidth ($\gtrsim 100$ MHz) with $\sim 10$ arcmin resolution is needed 
to perform a cosmological neutral hydrogen survey over a significant fraction of the sky. We have shown that 
a nearly packed interferometer array with few hundred receiver elements spread over an hectare or a hundred beam 
focal plane array with a $\sim \hspace{-1.5mm} 100  \, \mathrm{meter}$ primary reflector will have the required sensitivity to measure 
the 21 cm power spectrum. A method to compute the instrument response for interferometers  
has been developed and we have  computed the noise power spectrum for various telescope configurations.
The Galactic synchrotron and radio sources are a thousand time brighter than the redshifted 21 cm signal, 
making the measurement of this latter signal a major scientific and technical challenge. We have also studied  the performance of a simple foreground subtraction method through realistic models of the sky 
emissions in the GHz domain and simulation of interferometric observations. 
We have been able to show that the cosmological 21 cm signal from the LSS should be observable, but 
requires a very good knowledge of the instrument response. Our method has allowed us to define and 
compute the overall  {\it transfer function} or {\it response function} for the measurement of the 21 cm 
power spectrum. 
Finally, we have used the computed noise power spectrum and  $P(k)$ 
measurement response function to estimate 
the precision on the determination of Dark Energy parameters, for a 21 cm BAO survey. Such a radio survey 
could be carried using the current technology and would be competitive with the ongoing or planned 
optical surveys for dark energy,  with a fraction of their cost.
 

\bibliographystyle{aa}

\end{document}